\documentclass[twocolumn, amsmath, amssymb, aps]{revtex4-2}

\usepackage[dvipsnames]{xcolor}
\usepackage{graphicx}
\usepackage{enumitem}
\usepackage{siunitx}
\DeclareSIUnit\gauss{G}
\usepackage{booktabs}
\usepackage{multirow}
\usepackage{mathtools}

\usepackage[switch]{lineno}
\usepackage[utf8]{inputenc}

\newcommand\myshade{30}
\colorlet{mylinkcolor}{red}
\colorlet{mycitecolor}{orange}
\colorlet{myurlcolor}{orange}

\usepackage[%
bookmarks=true,
colorlinks,
linkcolor=mylinkcolor!\myshade!black,
urlcolor=myurlcolor!\myshade!black,
citecolor=mycitecolor!\myshade!black,
plainpages=false,
pdfpagelabels,
final,
breaklinks=true
]{hyperref}

\hypersetup{
	pdftitle={Momentum Exchange Interaction in a High-Finesse Cavity},
	pdfauthor={Thompson Lab},
	pdfkeywords={many-body interaction, entanglement, quantum}
}

\usepackage[normalem]{ulem}

\usepackage{physics}
\usepackage{braket}

\begin{document}
	\title{Cavity-Mediated Collective Momentum-Exchange Interactions}
	
	\author{Chengyi Luo, Haoqing Zhang}
	
	\author{Vanessa P.~W.~Koh, John D. Wilson, Anjun Chu,\\ Murray J. Holland, Ana Maria Rey}
	
	\author{James K.~Thompson}
	\email{jkt@jila.colorado.edu}
	\affiliation{JILA, NIST, and Department of Physics, University of Colorado, Boulder, CO, USA.}
	\date{\today}
	
	\begin{abstract}
		
		Quantum simulation and sensing hold great promise for providing new insights into nature, from understanding complex interacting systems to searching for undiscovered physics.  Large ensembles of laser-cooled atoms interacting via infinite-range photon mediated interactions are a powerful platform for both endeavours. Here, we realize for the first time momentum-exchange interactions in which atoms exchange their momentum states via collective emission and absorption of photons from a common cavity mode. The momentum-exchange interaction leads to an observed all-to-all Ising-like interaction in a matter-wave interferometer, which is useful for entanglement generation.  A many-body energy gap also emerges, effectively binding interferometer matter-wave packets together to suppress Doppler dephasing, akin to Mössbauer spectroscopy.  The tunable momentum-exchange interaction provides a new capability for quantum interaction-enhanced matter-wave interferometry and for realizing exotic behaviors including simulations of superconductors and dynamical gauge fields.

	\end{abstract}
	
	\maketitle

	Many-body  quantum states of laser-cooled atoms can be exquisitely controlled, making them powerful platforms for quantum simulation, metrology, and computing.  In particular, quantum sensing and metrology rely on understanding how to realize new forms of interactions between the atoms to achieve the next generation of ultra-precise quantum sensors and to emulate both  complex  quantum phases of matter  as well as non-equilibrium systems that are difficult to access in real materials.

	Optical cavities can be used to enhance the interaction of atoms with light in quantum many-body systems in which either the atomic internal \cite{leroux2010implementation,davis2019photon,norcia2018cavity,Thompson2020DynamicalPhaseTransition,sauerwein2022engineering}, motional \cite{black2003observation,baumann2010dicke,schuster2020supersolid,guo2021optical,kongkhambut2022observation}, or both \cite{kroeze2018spinor,Thompson2022SAI} degrees of freedom are coupled between different atoms.  In addition, the strong light-atom interaction has enabled the largest directly observed entanglement generation to date in any system \cite{cox2016deterministic,KasevichSqueezing2016}, with applications in quantum sensing with matter-wave interferometers \cite{Thompson2022SAI} and clocks \cite{VuleticSqueezedOpticalClock2020,robinson2022direct,VladanEntangledClockLifetime2010,KasevichSqueezing2016}.

\begin{figure*}[!hbt]
		\centering
		\includegraphics[width=0.95\textwidth]{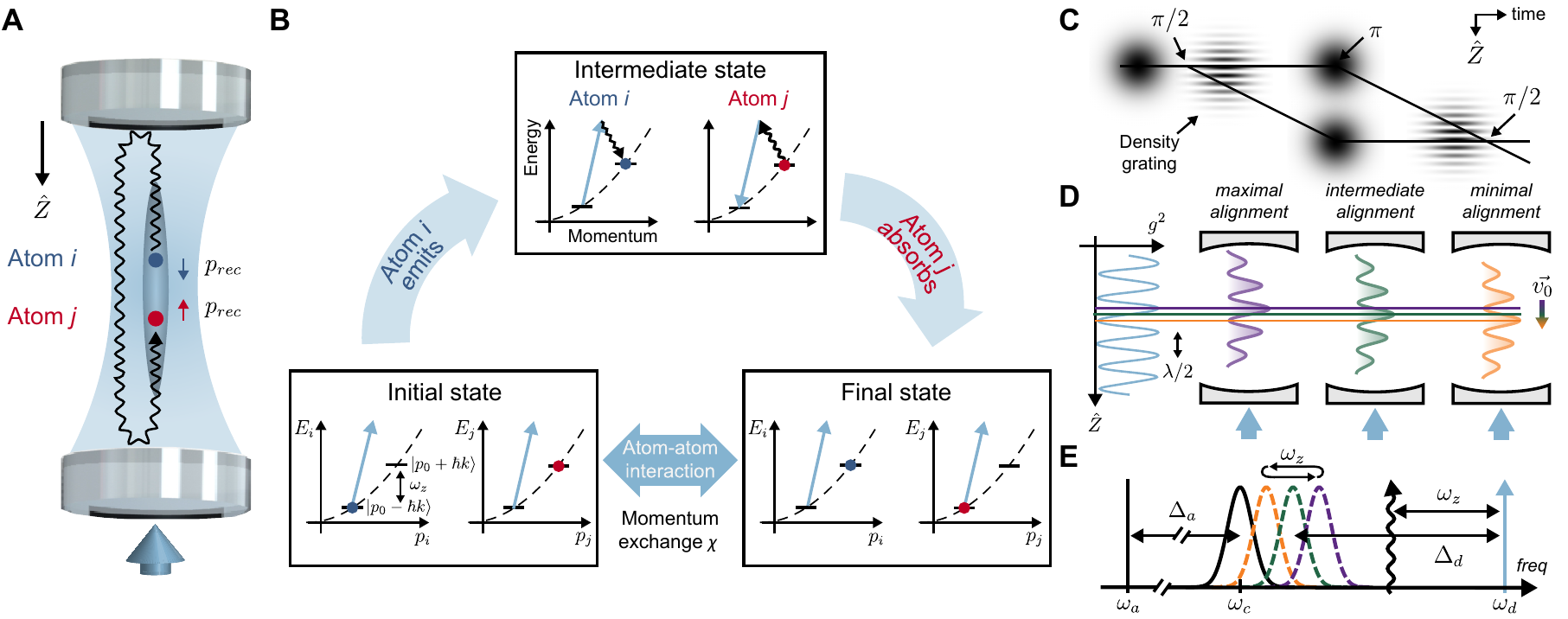}
		\caption{\textbf{Momentum exchange interaction.} 
			\textbf{(A)}  A momentum-exchange interaction is realized between atoms in different momentum states $\left(p_0\pm\hbar k\right)\hat{Z}$ by exchange of photons via a standing-wave optical cavity, illustrated for two particular atoms in red and blue. The dressing laser (light blue arrow) is injected into the cavity.
			\textbf{(B)} The energy versus momentum diagrams illustrate the  steps of the momentum-exchange between the example red and blue atoms. Eliminating the intermediate states leads to an effective momentum-exchange Hamiltonian involving only the atoms.  
			\textbf{(C)} Space-time diagram of the matter-wave interferometer. Bragg pulses are applied to manipulate atoms in superpositions of momentum states, causing the wave packets to separate in time with subsequent pulses reoverlapping the wave packets. When the wave packets overlap with each other, their interference forms a density grating along $\hat{Z}$.
			\textbf{(D)} As the atomic density grating moves, its spatial overlap with the standing-wave cavity mode (light blue on the left) varies, with three snapshots in time (purple, green and orange) shown on the right. 
			\textbf{(E)} Frequency diagram of the optical atomic transition frequency $\omega_a$ (black solid line), bare cavity frequency $\omega_c$ with no atoms in the cavity (black solid Lorentzian), and the atom-dressed cavity resonance frequency (dashed purple, green, orange Lorentzians) for the corresponding snapshots in time from (C). The average dressing laser detuning $\Delta_d$ is shown. The cavity is frequency modulated at $\omega_z$, leading to sidebands on the dressing laser at $\pm \omega_z$ (lower sideband shownas wiggly black line) that with the dressing laser couple the momentum states $p_0\pm\hbar k$ to realize the momentum-exchange. 
		}
		\label{fig1}
	\end{figure*}

	Here, we realize for the first time a cavity-mediated momentum-exchange interaction in a many-body system in which pairs of atoms exchange their momentum states, as shown in Fig.~\ref{fig1}A and B. The momentum exchange interaction arises from an atomic density grating creating sideband tones on an applied dressing laser, similar to as occurs in cavity opto-mechanical systems \cite{LehnertRegal2022Transduction,polzik2021membrane, StamperKurnOptoMech2016,Esslinger2008BECOptoMech}, as illustrated in Fig.~\ref{fig1}C to E. The momentum-exchange can be modeled  as an all-to-all pseudo-spin-exchange interaction, analogous to that observed for internal spin states \cite{norcia2018cavity,davis2019photon,Schleier-Smith2020GapProtection,Thompson2020DynamicalPhaseTransition,sauerwein2022engineering}. While previous  theoretical proposals have considered the generation of such momentum-exchange in a ring cavity, as well as extensions to two-mode squeezing involving additional spin degrees of freedom \cite{HollandSqueezedBraggAI,WilsonHollandTwoMode2022}, here we experimentally realize a momentum-exchange interaction in a standing wave cavity by exploiting the Doppler shift of the falling atoms.

	The observed momentum-exchange interaction allows for the realization of the collective XX-Heisenberg model, an iconic model in quantum magnetism  and superconductivity \cite{Marino2022, richardson1964exact,richardson1964pairing},  now generated  in a momentum-only basis of states with no internal atomic degrees of freedom involved, as compared to previous  \cite{norcia2018cavity,davis2019photon,Schleier-Smith2020GapProtection,Thompson2020DynamicalPhaseTransition} and contemporaneous work \cite{finger2023spin}. The exchange interaction manifests firstly as a magnetization-dependent global spin precession of the collective pseudo-spin Bloch vector, referred to as one-axis twisting (OAT). Secondly, it generates a many-body energy gap that realizes a collective recoil mechanism that suppresses dephasing due to Doppler broadening (i.e. single-particle dispersion), analogous to, but distinct from Mössbauer and Lamb-Dicke spectroscopy, which have been keys to realizing state-of-the-art quantum metrology and searches for new physics \cite{PoundRebka1960, Ye2022RedShift}. 
	
	This work also sets the stage for many-body simulation that goes beyond two-level systems by encoding degrees of freedom  in the larger ladder of momentum states \cite{An2018,Meir2018} as well as internal states \cite{Celi2014,Chalopin2020,Stuhl2015,Kolkowitz2017, Mancini2015}. The large number of synthetic dimensions in  combination with the long-range cavity mediated interactions,  open unique opportunities for the  realization of self-generated spin-orbit coupling \cite{Mivehvar2021,Chanda2021selforganized}, pair production \cite{davis2019photon,Ma2011,Meinert2016,Xu2018,lucke2011twin,gross2012spin, Parkins2017,periwal2021programmable}, long sought but never seen topological superfluids \cite{Kraus2013}, and dynamical gauge fields \cite{Schweizer2019,Gorg2019,Clark2018}.  In each case, the fragile correlations and intrinsic quantum properties  can be detected in both a non-destructive way and with beyond-mean-field sensitivity by the aid of the same atom-cavity interactions used to generate them \cite{cox2016deterministic,KasevichSqueezing2016, Thompson2022SAI}. Lastly, the generation of sideband tones may open a path to transduce excitations between matter waves and mesoscopic opto-mechanical systems \cite{LehnertRegal2022Transduction,polzik2021membrane} or back-action evading measurements of matter waves as proposed for spins \cite{shankar2019continuous}.

	In the experiment, $^{87}$Rb atoms are laser-cooled inside a two-mirror standing wave cavity that is vertically-oriented along $\ensuremath{\hat{Z}}$, see Fig.~\ref{fig1}A and \cite{Thompson2022SAI,CoxFallingSqueezing2016}. The atoms are allowed to fall along the cavity axis, guided by a blue detuned intracavity optical dipole trap.  A pair of laser beams with different frequencies are injected non-resonantly into the cavity in order to drive velocity-sensitive two-photon Raman transitions between ground hyperfine states (for state preparation and readout) or Bragg transitions that only change momentum states (for manipulating the superposition of momentum states.)
	
	The atoms are prepared in the ground hyperfine state $\ket{F=2,m_F=2}$ with momentum along the cavity axis $p_0-\hbar k$ and rms momentum spread $\sigma_p<0.1\hbar k$ where $\hbar$ is the reduced Planck constant, the wavenumber is $k=2\pi/\lambda$, and the wavelength is $\lambda = 780$~nm. As shown in Fig.~\ref{fig1}C, the Bragg lasers are then applied to place the atoms in a superposition of two wave packets with momenta centered on $p_0 \pm \hbar k$ with average momentum $p_0$ and separated by two photon recoil momenta $2 \hbar k$. Inserting additional Bragg pulses, we can realize a matter-wave interferometer, in which the atomic wave packets move apart and then reoverlap at later time. Just after the wave packet splitting and just before reoverlapping, the two portions of the wave packets interfere, leading to a spatially varying atomic density grating with periodicity $\lambda/2$ matching the periodicity of the standing-wave cavity mode.   
	
	As shown in Fig.~\ref{fig1}E, a cavity mode's frequency is detuned by $\Delta_{a}= 2\pi \times 500~\mathrm{MHz}$ to the blue of the D2 cycling transition $\ket{F=2,m_F=2} \rightarrow \ket{F'=3,m_{F'}=3}$.  A dressing laser with photon flux $|\alpha_d|^2$ (in unit of photons per second) drives the cavity at frequency $\omega_d$ that is typically within a $\mathrm{MHz}$ of the cavity resonance frequency. The input coupling of the cavity $\kappa_1$ is determined by the transmission of the input mirror. The detuning $\Delta_a$ is large compared to all other relevant frequency scales including the excited state decay rate $\Gamma=2 \pi \times 6$~MHz and the cavity power decay rate $\kappa = 2\pi \times 56(3)~$kHz. In this far-detuned limit, an atom at position $Z$ shifts the cavity resonance by $\frac{g_0^2}{\Delta_a} \cos^2 (k Z)$, where $g_0=2\pi \times 0.48~\mathrm{MHz}$ is the maximal Jaynes-Cummings atom-cavity coupling at a cavity anti-node \cite{ZilongsPRA}.
	
	\vspace{2mm}
	\noindent\textbf{Modulation sidebands.} As the atomic density grating moves along the cavity axis at velocity $v_0=p_0/m$, with $m$ the mass of $^{87}$Rb, the density grating goes from being aligned to misaligned with the cavity standing wave shown in Fig.~\ref{fig1}D from left to right. This leads to a modulation of the cavity resonance frequency at the two-photon Doppler frequency $\omega_z=2k v_0$ as shown in Fig.~\ref{fig1}E. The modulation of the cavity resonance frequency leads to optical modulation sidebands on the dressing laser inside the cavity at frequencies $\omega_d\pm\omega_z$, with the closer to resonance sideband shown in Fig.~\ref{fig1}E (black wiggly line), in a direct analogy to cavity opto-mechanical systems \cite{LehnertRegal2022Transduction,polzik2021membrane, StamperKurnOptoMech2016}. The modulation sidebands can also be understood as the Doppler-shifted reflection of the dressing laser from the moving matter-wave grating.

\begin{figure*}[ht]
		\centering
		\includegraphics[width=0.9\textwidth]{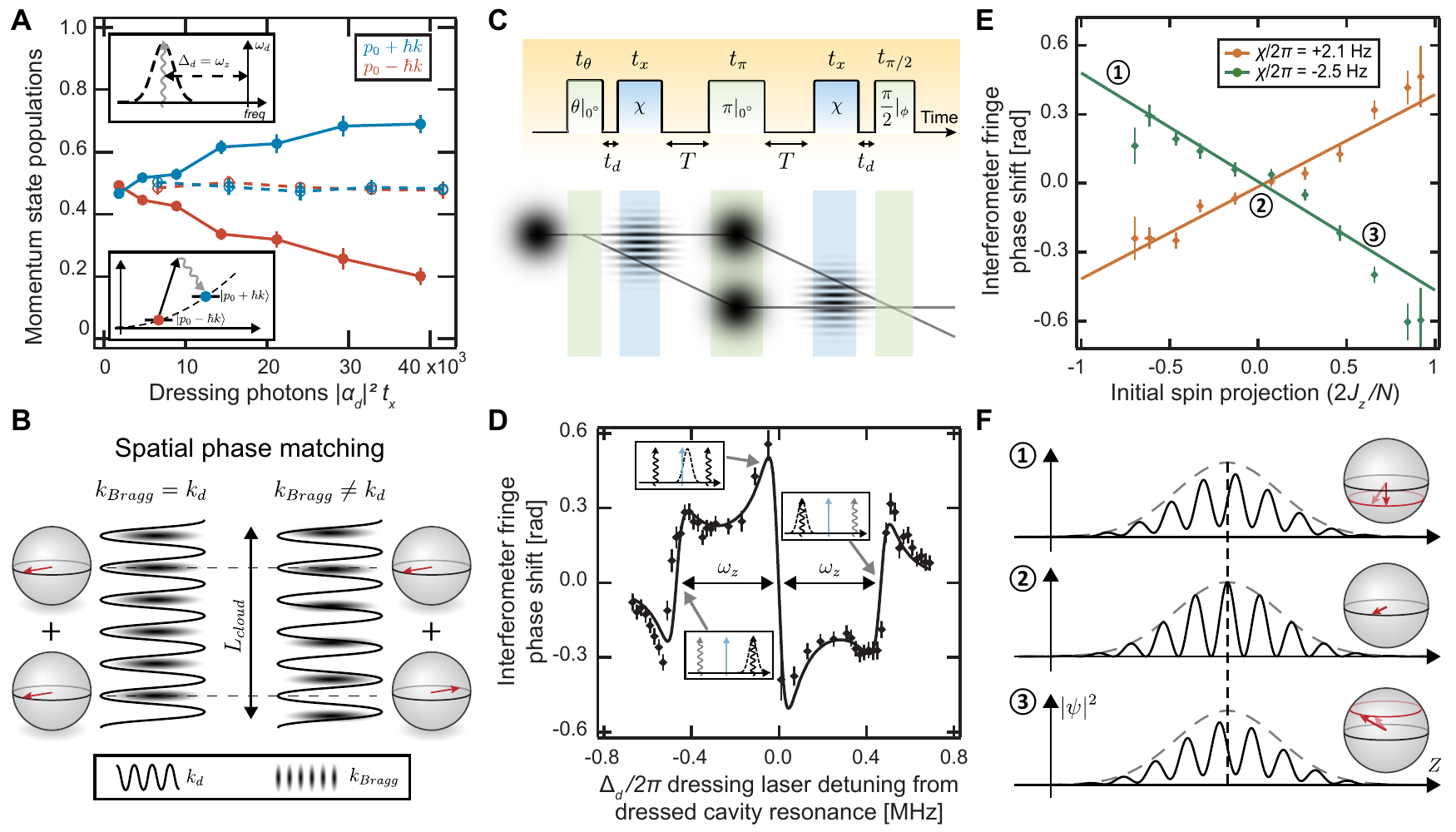}
		\caption{\textbf{Modulation sidebands and one-axis twisting dynamics.}
			\textbf{(A)} When a modulation sideband generated by the moving atomic density grating is tuned to resonance with the cavity (top inset), the light escapes the cavity and population is collectively or superradiantly transferred (bottom inset) between the two momentum states at $p_0\pm \hbar k$ (solid points and lines.) 
			\textbf{(B)} The system is well phase matched when the wave numbers of the Bragg laser (that generates the density grating) and the dressing laser (that drives the momentum-exchange interaction) closely match each other (on the left). A difference in wave number ($k_{Bragg}\neq k_d$) will lead to a spatially varying phase that eliminates the superradiance (on the right). 
			\textbf{(C)} Matter-wave interferometer sequence and space-time diagram for observing all-to-all Ising or One-Axis Twisting dynamics. The interferometer fringe amplitude and phase shift $\Delta \phi$ are measured by scanning the phase of the final rotation $\phi$. 
			\textbf{(D)} The observed phase shift $\Delta\phi$ of the interferometer fringe versus the dressing laser's detuning from the dressed cavity resonance, displaying the predicted (line) functional form of $\chi$ from Eq.~\ref{chiEq}.  The insets illustrate the relative alignment of the modulation sideband to the cavity resonance for three characteristic detunings.
			\textbf{(E)} The measured interferometer phase shift scales linearly with the initial spin projection $J_z = \langle \hat{J}_z \rangle$, while now holding $\Delta_d$ fixed.  The orange data points and fitted line is for $\chi/2\pi=+2.1$~Hz, and green for $\chi/2\pi=-2.5$~Hz). 
			\textbf{(F)} Visualizations of the phase shift $\Delta\phi$ in both the pseudo-spin picture (Bloch spheres) and in the atomic density grating picture for three characteristic points in (E).
		}
		\label{fig2}
	\end{figure*}

	We directly observe that a modulation sideband combined with the dressing laser form a Bragg coupling that drives collective population transfer from $p_0-\hbar k$ to $p_0+\hbar k$ as shown by the solid points and lines in Fig.~\ref{fig2}A.  This occurs when we tune a modulation sideband to be nominally on resonance with the dressed cavity by setting the dressing laser detuning from the average cavity resonance frequency (see Fig.~\ref{fig1}E) to $\Delta_d=\omega_z$. In this regime, the sideband light can escape from the cavity before being re-absorbed by the atoms, such that the collective population transfer can also be understood as a superradiant decay between momentum states. To confirm the collective nature of the decay, in a separate experiment, we prepare the initial superposition of states using an initial Bragg $\pi/2$-pulse $85$~GHz detuned from the dressing laser. The difference in wave numbers of the dressing laser $k_d$ and the Bragg laser $k_{Bragg}$ causes a slip in the spatial alignment of the cavity standing wave and the atomic density grating by a phase $2 \left| k_d-k_{Bragg} \right| L_{cloud}=3.5$~radians across the axial extent $L_{cloud}=1$~mm of the atomic cloud (Fig.~\ref{fig2}B). In this case, we observe no superradiant transfer of population in Fig.~\ref{fig2}A (open circles and dashed lines).

	We now realize the momentum-exchange interaction by tuning the dressing laser so that the modulation sidebands are far from resonance with the cavity. In this limit, photons emitted at the sideband frequencies are more likely to be re-absorbed by the atoms than to escape from the cavity. This process of emitting and absorbing sideband photons leads to a momentum-exchange as illustrated in Fig.~\ref{fig1}A and B. 
	
	\vspace{2mm}
	\noindent\textbf{Effective Hamiltonian.} To model the momentum exchange process, we begin by defining $\hat{\psi}^\dagger \left(p\right)$ and $\hat{\psi}\left(p\right)$ as creation and annihilation field operators of an atom with momentum $p$ which are related to creation and annihilation operators in position space by $\hat{\psi}(Z) = \int \hat{\psi}(p) e^{ipZ/\hbar}dp$. Because the wave packets centered at $p_0\pm \hbar k$ have a narrow momentum spread $\hbar k\gg\sigma_p$, we define $\hat{\psi}_{\uparrow }(p) = \hat{\psi}(p+p_0 + \hbar k)$, $\hat{\psi}_{\downarrow }(p) = \hat{\psi}(p+p_0 -\hbar k)$ operators that annihilate atoms at momentum $p+p_0\pm\hbar k$ within a momentum range $p\in [-\hbar k,+\hbar k]$. Doing this will support understanding in terms of both wave packets and an effective pseudo-spin language.
	
	We divide the differential kinetic energy between the two momentum states $p+p_0 \pm\hbar k$ into two terms: a homogeneous or common kinetic energy difference $\hat{H}_z(p)=\frac{\hbar\omega_z}{2} \left[\hat{\psi}_{\uparrow }^\dagger(p) \hat{\psi}_{\uparrow }(p) - \hat{\psi}_{\downarrow }^\dagger (p)\hat{\psi}_{\downarrow }(p)\right]$ and an inhomogeneous contribution $\hat{H}_{in}(p) = \frac{\hbar \omega_{in}(p)}{2} \left[\hat{\psi}_{\uparrow }^\dagger (p) \hat{\psi}_{\uparrow }(p) - \hat{\psi}_{\downarrow }^\dagger(p) \hat{\psi}_{\downarrow }(p)\right]$ with $\omega_{in}(p)=2k p/m$.

	We can adiabatically eliminate the cavity fields using second order perturbation theory (see Fig.~\ref{fig1}B), and in the perturbative limit $|\Delta_d\pm \omega_z|\gg \sqrt{N} \left| \frac{\alpha_d \sqrt{\kappa_1}}{\Delta_d + i \kappa/2} \right| \frac{g_0^2} {4\Delta_a} $, we obtain an effective atomic-only momentum-exchange Hamiltonian 
	\begin{equation}
	\begin{split}
	\hat{H}_{mx}= &\iint_{-\hbar k}^{\hbar k} \Bigr[ \hbar \chi_+  \hat{\psi}_{\uparrow }^\dagger(p) \hat{\psi}_{\downarrow }(p) \hat{\psi}_{\downarrow }^\dagger(q) \hat{\psi}_{\uparrow }(q)   \\ 
	& + \hbar  \chi_-   \hat{\psi}_{\downarrow }^\dagger (p) \hat{\psi}_{\uparrow }(p) \hat{\psi}_{\uparrow }^\dagger (q) \hat{\psi}_{\downarrow }(q) \Bigr] \,dp\,dq
	\end{split}
	\end{equation} 
	with the total Hamiltonian $\hat{H}=\hat{H}_{mx} + \int_{-\hbar k}^{\hbar k} \hat{H}_{in} (p) \, dp + \int_{-\hbar k}^{\hbar k} \hat{H}_{z} (p) \, dp$. The momentum exchange couplings are given by
	\begin{equation}
	\chi_\pm = \left( \frac{g_0^2}{4\Delta_a}\right)^2  {\frac{|\alpha_d|^2 \kappa_1}{\Delta_d^2+\kappa^2/4} }  {\frac{\Delta_d\pm\omega_z}{(\Delta_d\pm\omega_z)^2+\kappa^2/4}} \;,
	\label{chiEq}
	\end{equation}
	\noindent where we have included finite cavity damping via appropriate Lindblad operators (see supplement.)

	To map this to a pseudo-spin model, we define ladder operators $\hat{j}_+(p)=\hat{\psi}^\dagger_{\uparrow }(p) \hat{\psi}_{\downarrow } (p)$, $\hat{j}_-(p)=\hat{\psi}^\dagger_{\downarrow }(p) \hat{\psi}_{\uparrow } (p)$ and spin projection operators $\hat{j}_x (p)= \frac{1}{2}\left[\hat{j}_+(p) + \hat{j}_-(p)\right]$, $\hat{j}_y(p)=\frac{1}{2 i}\left[\hat{j}_+(p) - \hat{j}_-(p)\right]$ and $\hat{j}_z(p)=\frac{1}{2} \left[\hat{\psi}^\dagger_{\uparrow }(p) \hat{\psi}_{\uparrow }(p) - \hat{\psi}^\dagger_{\downarrow }(p) \hat{\psi}_{\downarrow }(p)\right]$.  Integrating over all momentum states, we can then define collective operators $\hat{J}_\alpha = \int_{-\hbar k}^{\hbar k} \hat{j}_\alpha (p)\,dp$ where $\alpha \in [x,y,z,+,-]$. The momentum-exchange Hamiltonian $\hat{H}_{mx}$ is then equivalent to an effective spin-exchange Hamiltonian
	$ \hat{H}_{sx}  = \chi_+ \hat{J}_+ \hat{J}_- + \chi_- \hat{J}_- \hat{J}_+$. This can be viewed as a collective XX-Heisenberg or Richardson-Gaudin integrable model where the non-local spin-spin couplings $\chi$ compete with an inhomogeneous axial field---a model often used in quantum magnetism  and superconductivity via the spin Anderson mapping \cite{Marino2022, richardson1964exact,richardson1964pairing}. We also note that the standing-wave cavity mode's spatial intensity variation $\cos^2 k Z$ produces additional terms $J_+^2$ and $J_-^2$ that we can neglect due to the same perturbative limit because these terms do not conserve energy between the initial and final states shown in Fig.~\ref{fig1}B (see supplement and \cite{borregaard2017one}.)

	\vspace{2mm}
	\noindent\textbf{One-axis twisting dynamics.} The exchange Hamiltonian can be re-written as $\hat {H}_{sx} \approx \chi \left( \hat{J}^2-\hat{J}_z^2 \right)$ with $\chi=\chi_+ +\chi_-$, ignoring single-particle terms. At the mean field level, the one-axis twisting Hamiltonian $\chi \hat{J}_z^2\approx 2 \chi \langle \hat{J}_z\rangle \hat{J}_z $  induces a rotation of the collective Bloch vector about the $z$ direction at a constant  frequency,  $2 \chi \langle \hat{J}_z\rangle$, that depends on the initial momentum population difference  $\langle \hat{J}_z\rangle$,  which is conserved by the Hamiltonian $\hat{H}_{sx}$. In the equivalent matter-wave picture, the azimuthal phase, $\Delta \phi = 2\chi \langle \hat{J}_z\rangle t_x$ accumulated  when the exchange interaction is applied for a time $t_x$, appears as a shift of the spatial interference fringe between the two wave packets, see Fig.~\ref{fig2}F. 
	
	To observe this phase shift, we run a matter-wave interferometer sequence (see Fig.~\ref{fig2}C) beginning with a Bragg $\pi/4$-pulse lasting 15~$\mu$s that prepares the atoms with population difference $\frac{\braket{ \hat{J}_z }}{N/2}\approx -0.7$. After waiting a delay time $t_{d}=25~\mu$s, we apply the dressing laser to create the exchange interaction for $t_{x}=25~\mu$s. To re-overlap the wave packets or equivalently undo the inhomogeneity from $\hat{H}_{in}$, we then apply a Bragg $\pi$-pulse, and apply the dressing laser again before applying a final Bragg $\pi/2$-pulse with various phase $\phi$. The final $\pi/2$-pulse maps the phase shift $\Delta\phi$ into a change in $\langle \hat{J}_z \rangle$. We measure the population in each momentum state by using velocity-sensitive Raman $\pi$-pulses and cavity-assisted quantum non-demolition measurements (see \cite{Thompson2022SAI} and supplement.) We repeat the experiment while scanning the phase of the final $\pi/2$-pulse. The phase shift $\Delta\phi$ is then determined from the phase of the observed fringe $\braket{ \hat{J}_z }$ versus $\phi$.
	
	The momentum-exchange coupling of Eq.\eqref{chiEq} predicts a triple-dispersive structure as the detuning of the dressing laser $\Delta_d$ varies. We observe this predicted structure in Fig.~\ref{fig2}D by measuring the induced phase shift $\Delta\phi$ as we vary the dressing laser detuning $\Delta_d$ from the dressed cavity resonance. In this data, the incident dressing laser power ($350~\mathrm{photons}/\mu$s) is held fixed. The two outer dispersive features arise as the two sideband frequencies at $\pm \omega_z$  pass through resonance with the cavity as shown in the insets. The dispersive feature near $\Delta_d=0$ arises from the carrier passing through resonance with the cavity. At $\Delta_d=0$, the exchange interaction parameters are $\chi_+\approx-\chi_-$ leading to a cancellation of the total exchange interaction ($\chi\approx 0$). 
	
	\begin{figure*}[!t]
		\centering
		\includegraphics[width=0.85\textwidth]{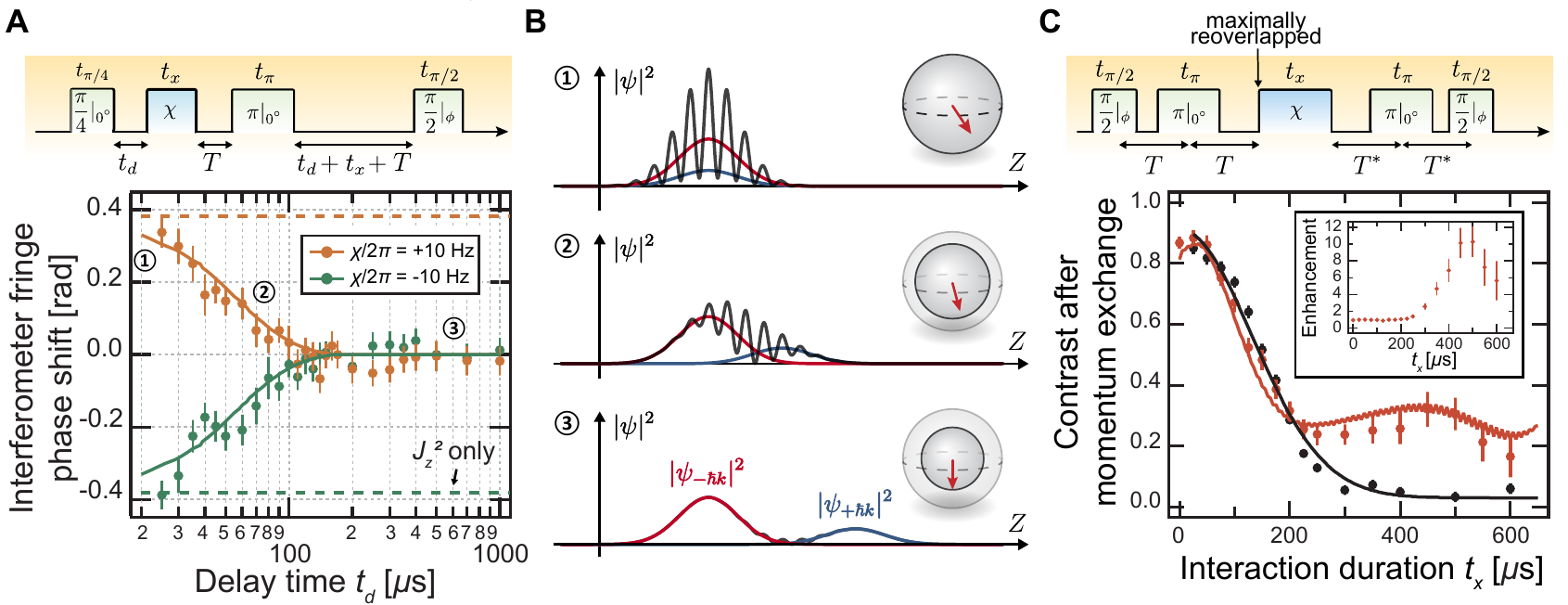}
		\caption{\textbf{Decoherence and energy gap protection.} 
			\textbf{(A)} The measured OAT phase shift goes to zero if the momentum-exchange interaction is applied after a delay time $t_d$ during which the atomic wave packets separate, measured for both positive and negative $\chi$, simulations with momentum-exchange interactions solid lines.  For comparison, the dashed lines are the simulated result with a pure OAT $\hat{J}_z^2$ Hamiltonian instead of the full momentum-exchange Hamiltonian.
			\textbf{(B)} Illustration of the wave packet separation and pseudo-spin representation at characteristic points in (A). As the wave packets separate, the atomic density grating disappears and the modulation sidebands that create the momentum-exchange interaction are no longer generated. Corresponding pseudo-spin Bloch spheres are shown.
			\textbf{(C)} Using the sequence in the top panel, the bottom panel shows the contrasts of the interferometer fringe measured at the end of the exchange interaction period with $\chi=0$ (black), $\chi/2\pi=6~$Hz (red). The ratio between the two (inset) displays significant gap protection of coherence due to the momentum-exchange's $\hat{J}^2$ contribution. The simulated results (solid lines) show good agreement with the data. 
		}  
		\label{fig3}
	\end{figure*}
	
	The phase shift $\Delta\phi$ is expected to scale linearly with $\braket{\hat{J}_z}$. We observe this by replacing the initial $\pi/4$-pulse with variable-length pulses to vary $\langle \hat{J}_z \rangle$ while holding $\Delta_d$ fixed instead. For the orange data in Fig.~\ref{fig2}E, the frequency of the relevant sideband is higher than the cavity resonance frequency leading to a measured $\chi/2\pi=+2.1$~Hz. For the green data, we retune the detuning $\Delta_d$ so that the relevant sideband frequency is lower than the cavity resonance frequency leading to a measured $\chi/2\pi=-2.5$~Hz. We observe a linear phase shift $\Delta\phi$ for $\chi>0$ and $\chi<0$ with opposite slopes as expected. 
	
	We observe that the size of the phase shift $\Delta\phi$ decreases if the wave packets are allowed to separate for a time $t_d$ before applying the dressing laser for time $t_x=25~\mu$s to induce the momentum-exchange interaction. Fig.~\ref{fig3}A (top) shows the pulse sequence used to measure this decay of the phase shift (bottom) for both positive and negative $\chi$ (orange and green points). For comparison, the solid lines indicate the predicted phase shifts for the full momentum-exchange Hamiltonian while the dashed lines indicate the predicted phase shift for an OAT Hamiltonian $-\chi\hat{J}^2_z$. The wave packet separation or equivalently the inhomogeneity $\hat{H}_{in}$ would not affect a pure OAT Hamiltonian as was the case in \cite{Thompson2022SAI}, whereas the phase shift is decreased by dephasing for exchange interaction as was observed in a spin system \cite{norcia2018cavity}. The wave packet separation leads to dephasing or shortening of the Bloch vector as visualized in Fig.~\ref{fig3}B. As the wave packets separate, the corresponding collective Bloch vectors are shortened while the projection $\langle \hat{J}_z \rangle$ is conserved. 
	
	
	\noindent\textbf{Gap protection: binding wave packets together.} 
	The additional non-linear term $\hat{J}^2$ in the momentum-exchange Hamiltonian gives rise to a many-body energy gap between states of higher symmetry (large $J$) and lower symmetry (smaller $J$) \cite{norcia2018cavity,Schleier-Smith2020GapProtection}. To explore how matter-wave coherence is protected by the gap, we run a Mach-Zehnder interferometer as shown in Fig.~\ref{fig3}C (top) in which we apply the dressing laser for a time $t_x$ starting at the point of maximum reoverlap of the wave packets (with $T=70~\mu$s). The coherence at the end of the dressing laser application is estimated from the amplitude of the interferometer fringe using an appropriately timed $\pi$-pulse and a final $\pi/2$-pulse shown (with $T^*=70~\mu$s.)  To account for the atomic loss resulting from free-space scattering and superradiance into higher momentum states, the contrast is calculated by normalizing the fitted fringe amplitude to the residual population in the two momentum states $p_0\pm \hbar k$. The actual coherence of the system is higher due to the finite possibility of under-estimating the number of atoms that underwent free-space scattering. In Fig.~\ref{fig3}C, the experiment is performed with the dressing laser off ($\chi=0$, black points and fitted black curve) and the dressing laser on ($\chi/2\pi=6$~Hz, red points and theory curve), for which we observe appreciable fringe contrast survives out to $600~\mu$s.  In Fig.~\ref{fig3}C inset, one sees that the momentum-exchange enhances the contrast by as much as a factor of 10(2).

	\begin{figure*}[!t]
		\centering
		\includegraphics[width=\textwidth]{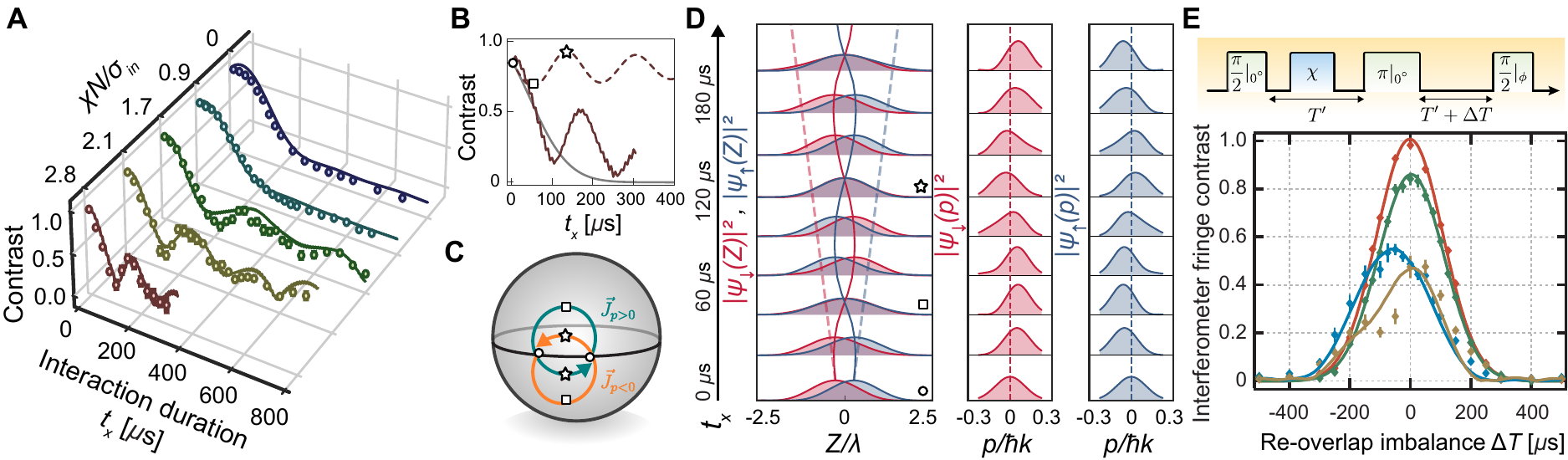}
		\caption{\textbf{Binding wave packets together.}   
			\textbf{(A)}  We run a similar sequence to that of Fig.~\ref{fig3}C except with an additional $40~\mu$s delay after wave packet overlap before application of the dressing laser for a variable time $t_x$.  We see that as the interaction strength is increased relative to the rms inhomogeneous broadening $\sigma_{in}=2\pi\times2$~kHz, there is a transition in the dynamics for $N \chi/\sigma_{in}>0.9$. Strikingly, there are also clear oscillations that were only hinted at in Fig.~\ref{fig3}C. The lines are theory predictions.
			\textbf{(B)} The theory prediction with residual superradiance on (solid) and turned off (dashed) are shown with three example points in the oscillations labeled, and with no interactions (grey).
			\textbf{(C)} The total length of the pseudo-spin Bloch vector $\vec{J}$ oscillates in time because the individual Bloch vectors oscillate as shown for $\vec{J}_{p>0}$ and $\vec{J}_{p<0}$ in green and orange respectively, with $\chi N /\sigma_{in} = 2.8$.
			\textbf{(D)} (left) In a co-moving frame, the wave packets oscillate in time about their average position in space (blue and red wave packets and centers solid lines, non-interacting system dashed lines.)  The momentum-space wave packets (right) also oscillate but with a $\pi/2$ phase shift in time relative to the position space wave packets, as would be the case for a harmonic oscillator. 
			\textbf{(E)} The interferometer contrast as a function of imbalance in the time from nominal perfect reoverlap of the wave packets with:  no momentum-exchange (red data and fit), momentum-exchange applied right after first $\pi/2$ at intermediate power (blue data and theory) and high power (brown data and theory), and momentum-exchange applied when wave packets are separated (green data and fit). 
		}  
		\label{fig4}
	\end{figure*}
	
	\vspace{2mm}
	
	In Fig.~\ref{fig3}C, the coherence undergoes a slight rise before ultimately falling.  This behavior can be accentuated by allowing the wave packets to undergo a small amount of separation for  $40~\mu$s before applying the momentum-exchange interactions for a duration $t_x$.  The observed interferometer contrast versus $t_x$  (see Fig.~\ref{fig4}A) is measured at different dressing laser powers to obtain different ratios of $\chi N$ to the rms inhomogeneity from $\hat{H}_{in}$ expressed as a frequency $\sigma_{in}$.  We observe a sharp transition in the dynamical behavior between $\chi N/\sigma_{in}$ = 0.9 and 1.7 with the emergence of oscillations of the contrast that extend to long times as $\chi N$ increases.  The oscillations become faster and have larger amplitudes at shorter times as $\chi N$ increases. This behavior is reasonably consistent with the overlayed theory simulations (colored traces in Fig.~\ref{fig4}A)  that  include finite superradiance and where only $\sigma_{in}$ is fit from the data with $\chi N =0$.

	The extension of coherence to longer times and the observed oscillations can be understood as the momentum-exchange interaction causing the wave packets to become bound to each other such that they no longer freely separate.  In Fig.~\ref{fig4}B, we show the simulated variation of the contrast versus time without superradiance, high lighting three example points that we explain using the simulated trajectories in Fig.~\ref{fig4}C for the collective pseudo-spin Bloch vectors evaluated for $p>0$ or $p<0$ with  $\vec{J}_{p>0} = \int_0^{\hbar k} \vec{j}(p)\,dp$ and $\vec{J}_{p<0} = \int_{-\hbar k}^{0} \vec{j}(p)\,dp$ where $\vec{j}(p)=\braket{ \hat{j}_x(p) \hat{x} + \hat{j}_y(p) \hat{y} + \hat{j}_z(p) \hat{z}}$. In Fig.~\ref{fig4}D, we also show the simulated results without superradiance for the individual wave packets in both momentum and position space.  In the pseudo-spin picture, the momentum-exchange causes the displayed vectors $\vec{J}_{p>0}$ and $\vec{J}_{p<0}$ to undergo orbits that oscillate symmetrically above and below the equator such that the total Bloch vector length oscillates in time.  In the wave packet picture, with no interactions, the wave packet centers would follow the diverging dashed lines.  With interactions, the wave packets oscillate in position with respect to each other, while also oscillating in their momentum $p$, as though the wave packets are now connected by a spring with characteristic frequency set by the exchange interaction strength $\chi N$ in the limit that $\chi N\gg\sigma_{in}$ and for small wave packet separation. If the wave packets are allowed to initially separate before the spring-like coupling is turned on, then the amplitude of the oscillations of the wave packet separations (and hence the contrast) will be larger as was observed in Fig.~\ref{fig4}A.
	
	To further explore this idea of wave packets becoming bound to each other, we run a Mach-Zehnder matter-wave interferometer with the sequence shown in Fig.~\ref{fig4}E (top).  If the exchange interaction is not applied (Fig.~\ref{fig4}E red points and fit), then the fringe contrast is maximized when the echo time difference is $\Delta T=0$ since this is when the wave packets have maximal reoverlap.  If the momentum exchange interaction is applied just after the first splitting pulse, we see that the point of maximum contrast is shifted to $\Delta T=-55~\mu$s (blue points and simulation), and becomes non-Gaussian (brown points and simulation) at even higher dressing laser power.  We rule out single-particle effects that might also shift the maximal reoverlap time by repeating the experiment, but with the dressing laser applied 1~ms after the first splitting pulse when the wave packets are not overlapped (green points and fit).  The fact that the delay is modified by $55~\mu$s rather than $t_x=25~\mu$s (as one might naively expect should one think of the wave packet separation as being frozen in place during the exchange interaction) arises from a $25~\mu$s delay between the end of the $\pi/2$-pulse and the beginning of the interaction, and the fact that the nature of the coupling of the wave packets is harmonic-oscillator-like.

	To emphasize the extreme unusualness of the binding of the wave packets, consider a \emph{gedanken} experiment in which all atoms start at rest and a single photon undergoes coherent two-photon absorption by the atoms with net momentum transfer $2\hbar k$, creating a symmetrized state with respect to which atom absorbed the photon.  Without exchange interactions, at long times one would observe a single atom emerge from the cloud with  velocity $v_{rec}=(2\hbar k)/m$ while all other atoms remain at their initial momentum. In contrast, with the momentum-exchange interaction, one would never observe a single atom to emerge with velocity $v_{rec}$. Instead the whole cloud of $N$ atoms would collectively recoil with velocity $v_{rec}/N$.
	
	This collective recoil is akin to Mössbauer spectroscopy (or Lamb-Dicke spectroscopy) in which atoms embedded in a crystal cause the whole crystal to recoil when the atoms absorb light.  However it is distinct in that here, firstly, the effect relies on a collective mechanism in which one fundamentally does not know \emph{which} atom absorbed photon momentum, and secondly, the atoms are in some sense continually swapping their quantum amplitudes for recoiling versus not recoiling via exchange of photons via the cavity.
	
	\vspace{2mm}
	\noindent\textbf{Summary.} 
	We have realized a cavity-mediated momentum-exchange interaction between different momentum states for the first time. By measuring the phase shift induced by this momentum exchange interaction, we observe the collective OAT dynamics, which paves the way for entanglement generation between momentum states and the study of beyond mean-field physics. We also directly observe an extension of the coherence time of the system which we identify with a collective recoil mechanism. The collective recoil mechanism still allows for the sensing of accelerations since the atoms still act as a phase memory of the optical Bragg pulses with which it interacts \cite{ThompsonGravWaveDetection2017}. This opens interesting new paths for Doppler-free spectroscopy and for matter-wave interferometers that do not rely on spin-echo like sequences and therefore would allow measurements of velocities rather than accelerations. 
	
	Finally, we note that the momentum-exchange Hamiltonian here is equivalent to the model Hamiltonian often used to describe BCS s-wave superconductors \cite{Marino2022,Lewis-Swan2021,smale2019observation}. From this perspective, the observed oscillations can be identified with Higgs oscillations following a quench of the exchange interaction strength. This  would enable quantum simulation  of the BEC-BCS crossover \cite{Parish,Randeria2014} akin to the one already observed in ultracold fermionic atoms but in   synthetic degrees  of freedom without the cooling  limitations or the three-body losses that  have limited the observation of a variety of exciting phenomena predicted to exist  in this system \cite{Marino2022}. Besides superfluidity, in the future  this system should offer the possibility to study pair-production and  self-generated interaction-induced  spin-orbit  coupling opening a path for quantum simulation of the Schwinger effect in  high energy physics \cite{Hauke_PhysRevX_2013,Kasper_PhysLettB_2016}, the Unruh thermal radiation in general relativity \cite{Hu_NatPhys_2019},  and thermofield double states  in the holographic correspondence relating a quantum-field theory to a gravitational theory in one higher dimension \cite{Chapman_SciPost_2019,Zhu_PNAS_2020}.

	\begin{acknowledgments}
		This material is based upon work supported by the U.S. Department of Energy, Office of Science, National Quantum Information Science Research Centers, Quantum Systems Accelerator. We acknowledge additional funding support from the National Science Foundation under Grant Numbers 1734006 (Physics Frontier Center) and  OMA-2016244 (QLCI), NIST, and DARPA/ARO W911NF-19-1-0210 and W911NF-16-1-0576, AFOSR grants FA9550-18-1-0319 and FA9550-19-1-0275.  We acknowledge helpful discussions with Eugene Polzik, Monika Schleier-Smith, and Athreya Shankar.
		
	\end{acknowledgments}

	\bibliography{Momentum_Exchange_Interaction}
	\clearpage
	\newpage

	\onecolumngrid
	\begin{center}
		\textbf{\Large{Supplemental Material}}
	\end{center}
	
	\makeatother
	\renewcommand{\theequation}{S\arabic{equation}}
	\setcounter{equation}{0}
	\renewcommand{\thetable}{S\arabic{table}}
	\setcounter{table}{0}
	\renewcommand{\thefigure}{S\arabic{figure}}
	\setcounter{figure}{0}
	\renewcommand{\bibnumfmt}[1]{[S#1]}
	\renewcommand{\citenumfont}[1]{S#1}
	
	\section{Derivation of momentum exchange Hamiltonian}
	
	\subsection{Original Hamiltonian in momentum basis}
	
	Here we start with the full atom-cavity Hamiltonian in the lab frame given as $\hat{H}_{\mathrm{lab}}=\hat{H}_{{\mathrm{atom}}}+\hat{H}_{{\mathrm{light}}}+\hat{H}_{{\mathrm{int}}}$ with the following second quantized form:
	\begin{align}
	\hat{H}_{{\mathrm{light}}} & =\hbar \sqrt{\kappa_1}\left(\alpha_{d} \hat{a}^{\dagger}e^{-i\omega_{d}t}+\alpha_{d}^{*} \hat{a}e^{i\omega_{d}t}\right)+\hbar\omega_{c}\hat{a}^{\dagger}\hat{a} , \\
	\hat{H}_{{\mathrm{atom}}} & = \sum_{\beta=g,e} \int \hat{\psi}_{\beta}^{\dagger}(Z) \frac{\hat{p}^{2}}{2m}\hat{\psi}_{\beta}(Z) \,dZ + \hbar \omega_{a} \int \hat{\psi}_{e}^{\dagger}(Z) \hat{\psi}_{e}(Z) \,dZ,  \\
	\hat{H}_{{\mathrm{int}}} & =\hbar g_0 \int \cos \left(kZ\right)\left[\hat{a}\hat{\psi}_{e}^{\dagger}(Z)\hat{\psi}_{g}(Z)+\hat{a}^{\dagger}\hat{\psi}_{g}^{\dagger}(Z)\hat{\psi}_{e}(Z)\right]\,dZ.
	\end{align}
	Here, $\alpha_d$ is the complex amplitude of the incident dressing laser with photon flux $|\alpha_d|^2$ in unit of photons per second, and $\kappa_1 = T_1 /\tau_{\mathrm{rnd}}$ is the input coupling of the cavity with $T_1$ the power transmission coefficient for the input mirror and $\tau_{\mathrm{rnd}}$ the optical round trip time in the cavity. $\hat{\psi}_{e(g)}^{\dagger}(Z)$ are the field creation operator for an atom at position $Z$ in excited state $\left|e\right\rangle \equiv \ket{F=3,m_F=3}$ and ground state $\left|g\right\rangle \equiv \ket{F=2,m_F=2}$, $\omega_{a}$ is the energy splitting between ground and excited state, $\delta_{d}=\omega_{d}-\omega_{c}$ is the detuning of the dressing laser from the empty cavity, and $\Delta_{a}=\omega_{c}-\omega_{a}$ is the detuning of the empty cavity from the atomic transition frequency. We rewrite the Hamiltonian in the rotating frame of the dressing laser at $\omega_d$, with the Hamiltonian used to construct the appropriate unitary $\hat{U}=\exp \left( i \hat{H}_r t\right)$ given by $\hat{H}_{r}=\hbar\omega_{d}\hat{a}^{\dagger}\hat{a}+\hbar\omega_{c}\int \hat{\psi}_{e}^{\dagger}(Z)\hat{\psi}_{e}(Z) \,dZ$.  In this frame, the Hamiltonian $\hat{H}_d=\hat{U}^\dagger \hat{H}_{\mathrm{lab}} \hat{U} + i\hbar \frac{\partial \hat{U}^\dagger}{\partial t}\hat{U}$ takes the form:
	
	\begin{equation}
	\begin{split}
	\hat{H}_d &= \hbar \sqrt{\kappa_1}\left(\alpha_{d} \hat{a}^{\dagger}+\alpha_{d}^{*} \hat{a}\right)-\hbar\delta_{d}\hat{a}^{\dagger}\hat{a}-\hbar\Delta_{a} \int  \hat{\psi}_{e}^{\dagger}(Z)\hat{\psi}_{e}(Z)\,dZ + \sum_{\tau=g,e} \int \hat{\psi}_{\tau}^{\dagger}(Z)\frac{\hat{p}^{2}}{2m}\hat{\psi}_{\tau}(Z) \,dZ \\
	&+ \hbar g_0 \int \cos (kZ) \left[\hat{a} \hat{\psi}_{e}^{\dagger}(Z) \hat{\psi}_{g}(Z)+\hat{a}^{\dagger} \hat{\psi}_{g}^{\dagger}(Z)\hat{\psi}_{e}(Z)\right] \,dZ.
	\end{split}
	\end{equation}

	Furthermore, we assume the excited state population and atomic spontaneous emission are negligible during the relevant time scales with $\Delta_{a}\gg g_0\sqrt{\left\langle \hat{a}^{\dagger}\hat{a}\right\rangle}$ and $\Delta_{a}\gg\gamma$ , where $\gamma$ is the spontaneous population decay rate from the optically excited state. In this limit, we can adiabatically eliminate the excited state $\left|e\right\rangle$, which leads to the effective Hamiltonian describing dispersive coupling between atoms in the ground state manifold and cavity:
	\begin{equation}
	\begin{split}
	\hat{H}_{\mathrm{disp}} & =\int \hat{\psi}^{\dagger}(Z)\left[\frac{\hat{p}^{2}}{2m}+\frac{\hbar g_{0}^{2}}{\Delta_{a}}\cos^{2}(k Z)\hat{a}^{\dagger}\hat{a}\right]\hat{\psi}(Z) \,dZ -\hbar \delta_{d}\hat{a}^{\dagger}\hat{a}+\hbar \sqrt{\kappa_1} \left( \alpha_{d}\hat{a}^{\dagger}+\alpha_{d}^{*}\hat{a} \right)\\
	& =\int \hat{\psi}^{\dagger}(Z)\left[\frac{\hat{p}^{2}}{2m}+\frac{ \hbar g_0^2}{4\Delta_a}\left(e^{i 2k Z}+e^{-i2k Z}\right)\hat{a}^{\dagger}\hat{a}\right]\hat{\psi}(Z)\,dZ-\hbar \Delta_{d} \hat{a}^{\dagger}\hat{a}+\hbar \sqrt{\kappa_1} \left( \alpha_{d}\hat{a}^{\dagger}+\alpha_{d}^{*}\hat{a} \right),
	\label{eqn:Hdisp}
	\end{split}
	\end{equation}
	with $\hat{\psi}(Z)\equiv\hat{\psi}_g (Z)$. This dispersive Hamiltonian tells us that an atom created at an anti-node of the cavity (i.e. $\cos^2(kZ)=1$) causing a shift of the cavity resonance by $\frac{g_0^2}{\Delta_a}$ as stated in the main text. For a uniform distribution of atoms along the cavity axis, the cavity’s dressed resonance frequency is on-average shifted by $\frac{Ng_0^{2}}{2\Delta_{a}}$. The detuning $\Delta_{d}=\delta_{d}-\frac{Ng_0^{2}}{2\Delta_{a}}$ can be physically interpreted then as the detuning of the dressing laser from the average dressed cavity resonance frequency.
	
	
	We then perform a Fourier transformation on the atomic field operator $\hat{\psi}(Z)$ to the momentum space as $ \hat{\psi}(Z)=\int_{-\infty}^{\infty} \hat{\psi}(p)e^{i p Z/\hbar} \,dp$, here $\hat{\psi}(p)$ annihilates an atom with momentum $p$. Due to the narrow momentum distribution after velocity selection with rms momentum spread $\sigma_p\ll \hbar k$, we only focus on momentum states within the range $[p_{0}-2\hbar k,p_{0}+2\hbar k]$. We further define $ \hat{\psi}_{\uparrow}(p) \equiv \hat{\psi}(p+p_{0}+\hbar k),\quad\hat{\psi}_{\downarrow}(p) \equiv \hat{\psi}(p+p_{0}-\hbar k), $ with $p\in[-\hbar k,\hbar k]$. The operator $\hat{\psi}_{\uparrow} (p)$ annihilates an atom with momentum $p+p_{0}+\hbar k_{c}$, while $\hat{\psi}_{\downarrow} (p)$ annihilates an atom
	with momentum $p+p_{0}-\hbar k$. 
	
	The original atomic field operator can now be expressed as
	\begin{equation}
	\begin{split}
	\hat{\psi}(Z) & \approx\int_{p_{0}-2\hbar k}^{p_{0}} \hat{\psi}(p)e^{ipZ/\hbar} \,dp+\int_{p_{0}}^{p_{0}+2\hbar k}\hat{\psi}(p)e^{ipZ/\hbar}\,dp\\
	&= e^{ip_{0}Z/\hbar}\int_{-\hbar k}^{+\hbar k} e^{ipZ/\hbar}\left(\hat{\psi}_{\downarrow}(p)e^{-i k Z}+\hat{\psi}_{\uparrow}(p)e^{i k Z} \right) \,dp . 
	\end{split}
	\end{equation}
	
	\noindent Substituting the approximate Fourier transform to momentum space into the dispersive Hamiltonian Eq.~\eqref{eqn:Hdisp}, we find
	\begin{equation}
	\begin{split}
	\hat{H} _{\mathrm{disp}} =& \int \hat{\psi}^{\dagger}(Z)\left[\frac{\hat{p}^{2}}{2m}+\frac{g_{0}^{2}}{4\Delta_{a}}(e^{2ik Z}+e^{-2ik Z})\hat{a}^{\dagger}\hat{a} \right]\hat{\psi}(Z)\,dZ+\hat{H}_{{\rm cavity}}\\
	\approx & \int_{-\hbar k}^{+\hbar k}\left[\hat{\psi}_{\downarrow}^{\dagger}(p)\hat{\psi}_{\downarrow}(p)\frac{(p+p_{0}-\hbar k)^{2}}{2m}+\hat{\psi}_{\uparrow}^{\dagger}(p)\hat{\psi}_{\uparrow}(p)\frac{(p+p_{0}+\hbar k)^{2}}{2m}\right] \,dp+\hat{H}_{{\rm cavity}}\\
	& +\frac{g_{0}^{2}}{4\Delta_{a}}\hat{a}^{\dagger}\hat{a}\int dZ\int_{-\hbar k}^{+\hbar k}\int_{-\hbar k}^{+\hbar k}\left[\hat{\psi}_{\downarrow}^{\dagger}(p)\hat{\psi}_{\uparrow}(q)e^{i(q-p)Z}+\hat{\psi}_{\uparrow}^{\dagger}(p)\hat{\psi}_{\downarrow}(q)e^{i(p-q)Z}\right]\,dp\,dq\\
	= & \frac{1}{2}\int_{-\hbar k}^{+\hbar k}\left[\hat{\psi}_{\uparrow}^{\dagger}(p)\hat{\psi}_{\uparrow}(p)-\hat{\psi}_{\downarrow}^{\dagger}(p) \hat{\psi}_{\downarrow}(p)\right] \left(\omega_{z}+\frac{2\hbar k p}{m}\right)\,dp+\hat{H}_{{\rm cavity}}\\
	& +\frac{g_0^{2}}{4\Delta_{a}}\hat{a}^{\dagger}\hat{a}\int_{-\hbar k}^{+\hbar k}\int_{-\hbar k}^{+\hbar k}\delta(p-q)\left[\hat{\psi}_{\downarrow}^{\dagger} (p)\hat{\psi}_{\uparrow} (q)+\hat{\psi}_{\uparrow}^{\dagger} (p)\hat{\psi}_{\downarrow} (q)\right] \,dp\,dq,
	\end{split}
	\end{equation}
	where the delta function that naturally emerges above arises from the physical fact that the two-photon processes can only couple momentum states separated by two photon momentum recoils $2 \hbar k$.
	
	We can break the above dispersive Hamiltonian into physically meaningful terms as
	\begin{equation}
	\hat{H}_{\mathrm{disp}} = \hat{H}_{{\rm z}}+\hat{H}_{{\rm in}}+\hat{H}_{{\rm cavity}}+\hat{H}_{{\rm atom-cavity}}.
	\label{eqn:Hdispsingle}
	\end{equation}
	
	\noindent The average kinetic energy difference $\hbar \omega_z = 2\hbar k p_0/m$ between the two momentum states $p_0+p\pm\hbar k$ (i.e. independent of $p$) is given by
	\begin{equation}
	\hat{H}_{{\rm z}} = \frac{\hbar\omega_{z}}{2} \int_{-\hbar k}^{+\hbar k} \left[\hat{\psi}_{\uparrow}^{\dagger}(p)\hat{\psi}_{\uparrow}(p)-\hat{\psi}_{\downarrow}^{\dagger}(p)\hat{\psi}_{\downarrow}(p)\right] \,dp.
	\end{equation}
	
	\noindent The inhomogeneous kinetic energy difference between the two momentum states $p_0+p\pm\hbar k$ (i.e. dependent on $p$)  is captured by 
	\begin{equation}
	\hat{H}_{{\rm in}} =\frac{\hbar k}{m}\int_{-\hbar k}^{+\hbar k} p \left[\hat{\psi}_{\uparrow}^{\dagger}(p)\hat{\psi}_{\uparrow }(p)-\hat{\psi}_{\downarrow}^{\dagger}(p)\hat{\psi}_{\downarrow}(p)\right] \,dp.
	\end{equation}
	
	\noindent The driven cavity is described by 
	\begin{equation}
	\hat{H}_{{\rm cavity}}  =-\hbar \Delta_d \hat{a}^{\dagger}\hat{a}+\hbar \sqrt{\kappa_1} \left( \alpha_{d} \hat{a}^{\dagger}+\alpha_{d}^{*}\hat{a} \right).
	\end{equation}
	
	\noindent The remaining atom-cavity coupling is described by
	\begin{equation}
	\begin{split}
	\hat{H}_{{\rm atom-cavity}}= \hat{a}^{\dagger}\hat{a} \, \frac{\hbar g_{0}^{2}}{4\Delta_{a}}\int_{-\hbar k}^{+\hbar k} \left[\hat{\psi}_{\downarrow}^{\dagger}(p)\hat{\psi}_{\uparrow}(p)+\hat{\psi}_{\uparrow}^{\dagger}(p)\hat{\psi}_{\downarrow}(p)\right]\,dp.
	\end{split}
	\end{equation}
	
	\noindent In the following, we will describe how to map the above Hamiltonian to a pseudo-spin model and show how to eliminate the cavity field and arrive at an effective atom-only interaction.

	\subsection{Mapping from momentum states to pseudo-spins}
	Using a Schwinger-boson representation, we can define pseudo-spin operators based on the two momentum states at $p+p_0\pm\hbar k$. First we define ladder operators 
	\begin{equation}
	\hat{j}_+= \hat{\psi}^\dagger_{\uparrow}(p) \hat{\psi}_{\downarrow}(p), 
	\quad \hat{j}_-=\hat{\psi}^\dagger_{\downarrow}(p) \hat{\psi}_{\uparrow}(p) ,
	\end{equation}
	from which we can also construct the spin angular momentum operators
	\begin{equation}
	\begin{split}
	\hat{j}_x(p) &= \frac{1}{2}\left[\hat{j}_+(p) + \hat{j}_-(p)\right]  , \\
	\hat{j}_y(p) &=\frac{1}{2 i }\left[\hat{j}_+(p) - \hat{j}_-(p)\right]  , \\
	\hat{j}_z(p) &=\frac{1}{2} \left[\hat{\psi}^\dagger_{\uparrow}(p) \hat{\psi}_{\uparrow}(p) - \hat{\psi}^\dagger_{\downarrow }(p) \hat{\psi}_{\downarrow}(p)\right],
	\end{split}
	\end{equation}
	
	\noindent where the operators satisfy the commutation relation $\left[ \hat{j}_x (p), \hat{j}_y (p)\right]= i \epsilon_{xyz} \,\hat{j}_z(p)$ plus all permutations of x, y, z, and where $\epsilon_{xyz}$ is the Levi-Civita symbol. Summing over all momentum states, we then define the collective spin operators
	\begin{equation}
	\hat{J}_\alpha = \int_{-\hbar k}^{\hbar k} \hat{j} (p) \, dp , \quad \alpha\in[x,y,z,+,-].
	\label{eqn:BigJ}
	\end{equation}
	
	\noindent Using only spin operators, we can now express the dispersive Hamiltonian Eq.~\eqref{eqn:Hdispsingle} in the pseudo-spin basis
	\begin{equation}
	\hat{H}_{\mathrm{disp}} = \hbar \omega_z \hat{J}_z+\frac{\hbar g_0^2}{4 \Delta_a} \hat{a}^{\dagger} \hat{a}(\hat{J}_{+}+\hat{J}_{-})-\hbar\Delta_d \hat{a}^{\dagger} \hat{a}+\hbar \sqrt{\kappa_1} \left( \alpha_{d}\hat{a}^{\dagger}+\alpha_{d}^{*}\hat{a} \right) + \hat{H}_{\mathrm{in}}. 
	\label{eqn:atomcavity}
	\end{equation} 
	\noindent In this pseudo-spin basis, we also re-write the inhomogeneous Hamiltonian as
	\begin{equation}
	\hat{H}_{\mathrm{in}} = \frac{2\hbar k}{m} \int_{-\hbar k}^{\hbar k} p\, \hat{j}_z (p) \, dp.
	\end{equation}
	
	\noindent In this pseudo-spin picture, the inhomogeneous kinetic energy can be thought of as an effective inhomogeneous magnetic field that shifts the energy difference between spin up and spin down with a linear dependence on $p$. 
	
	The time evolution of the density matrix describing the atom-cavity coupled system is given by the following Master equation:
	\begin{equation}
	\begin{split}
	\frac{d}{d t} \rho=-\frac{i}{\hbar}[\hat{H}_{\mathrm{disp}}, \rho]+\hat{L} \rho \hat{L}^{\dagger}-\frac{1}{2}\{\hat{L}^{\dagger} \hat{L}, \rho\}, 
	\end{split}
	\end{equation}
	\noindent where the jump operator describing the cavity power decay at rate $\kappa$ is given by
	\begin{equation}
	\hat{L} = \sqrt{\kappa} \hat{a}.
	\end{equation}
	
	\subsection{Effective Hamiltonian from second-order perturbation theory}
	In order to reach an effective atom-atom interaction from the Hamiltonian Eq.~\eqref{eqn:atomcavity}, we eliminate the cavity field with second-order perturbation theory. First, we express the cavity operator $\hat{a}$ as a cavity operator $\hat{b}$ that captures the cavity field dynamics around an average field described by a time-independent complex number
	\begin{equation}
	\hat{a} = \alpha_0 + \hat{b} ,
	\label{eqn:cavityfieldexpand}
	\end{equation}
	\begin{equation}
	\alpha_0 = \frac{\alpha_{d} \sqrt{\kappa_1}}{\Delta_d+i\kappa/2} ,
	\label{eqn:alpha0}
	\end{equation}
	
	\noindent where the complex number $\alpha_0$ in unit of $\sqrt{\mathrm{photons}}$ is the dressing laser’s field amplitude that builds up inside the cavity at a dressing laser detuning $\Delta_d$. Substituting Eq.~\eqref{eqn:cavityfieldexpand} back into the dispersive Hamiltonian Eq.~\eqref{eqn:atomcavity}, the Hamiltonian and jump operators become
	\begin{equation}
	\begin{split}
	\hat{H}_{\alpha_0} &= \hat{H}_0 + \hat{V} , \\
	\hat{H}_0 &= \hbar \omega_z \hat{J} _z - \hbar\Delta_d \hat{b}^\dagger \hat{b} , \\
	\hat{V} &= \frac{\hbar g_0^2}{4\Delta_a}  \left( \hat{J}_+ + \hat{J}_- \right) \left( \alpha_0 \hat{b}^\dagger +  \alpha_0^* \hat{b} \right) , \\
	\hat{L} &= \sqrt{\kappa} \hat{b} . 
	\end{split}
	\end{equation}
	
	\noindent Importantly, the interaction $\hat{V}$ can be interpreted as a field displacement operator acting on the cavity, but the amplitude of the displacement depends on the atomic degree of freedom $\hat{J}_+ +\hat{J}_-$. The term $\hat{J}_+ +\hat{J}_-$ will oscillate at frequency $\omega_z$ due to the $\hbar \omega_z \hat{J}_z$ term in $\hat{H}_0$. This term leads to the generation of modulation sidebands on the dressing laser light inside the cavity at absolute optical frequency $\omega_d \pm \omega_z$. In this picture, this is the origin of the modulation sidebands discussed in the main text. Conceptually, the sidebands that are generated represent the first-order perturbative modification of the cavity field about the zeroth order response $\alpha_0$.
	
	\begin{figure*}[!ht]
		\centering
		\includegraphics[width=0.7\textwidth]{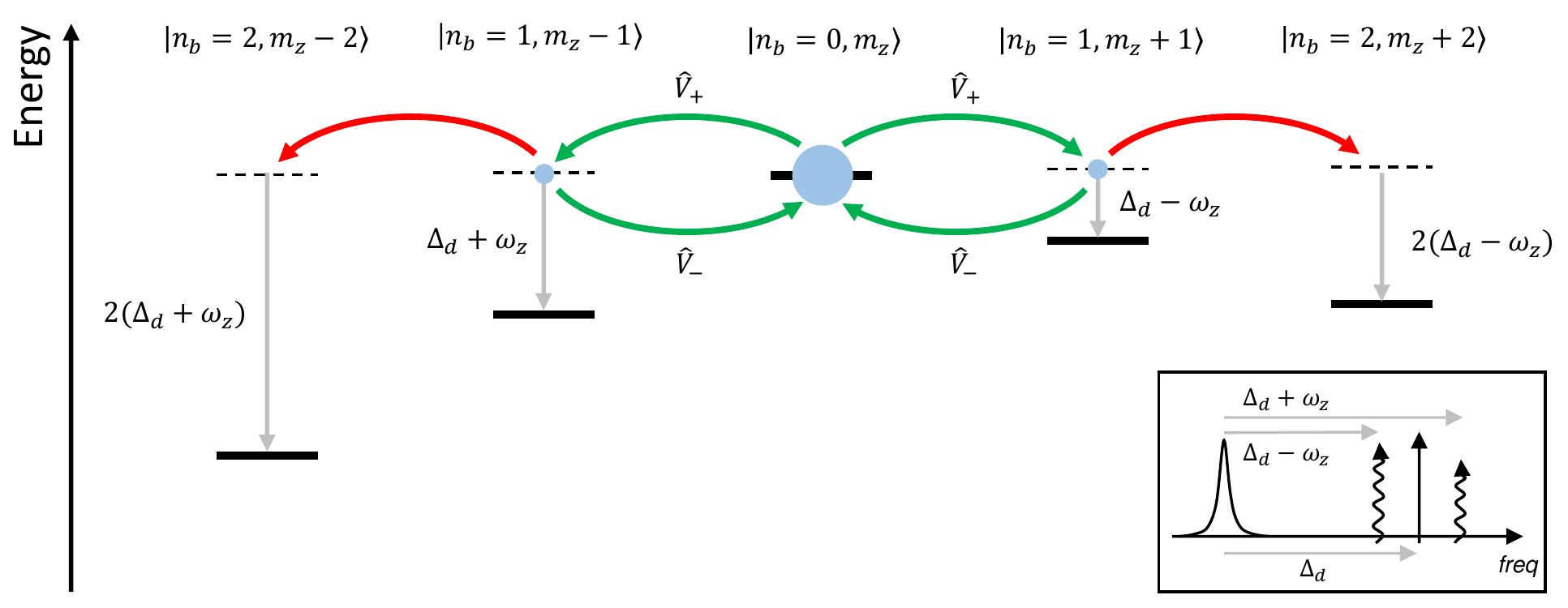}
		\caption{\textbf{Second order perturbation theory.} Relative energy level diagrams in the rotating frame of the dressing laser when applied at detuning $\Delta_d$ from the dressed cavity. In the second order perturbation theory, the initial states are coupled to the singly excited states and back with the processes labeled as green arrows. Due the the large energy offset $2\left(\Delta_d \pm \omega_z \right)$, the processes coupling doubly excited states (red arrows) are energetically forbidden. 
		}  
		\label{fig-s1}
	\end{figure*}
	
	To arrive at an atom-only Hamiltonian, we now go to the second order in perturbation theory by assuming that the $\left|\left\langle \hat{b}\right\rangle\right|\ll 1$  or equivalently the total amplitude of the generated sidebands is small, which is true if $|\Delta_d\pm\omega_z|\gg \sqrt{N} |\alpha_0| g_0^2 /4\Delta_a$. In this limit, we can use $\hat{V}$ as a perturbation on $\hat{H}_0$ and adiabatically eliminate the photon excitation to obtain an effective Hamiltonian in the zero photon subspace. We consider the basis $\ket{n_b, m_z}$, where $n_b$ is the photon number such that $\hat{b}^\dagger \hat{b} \ket{n_b, m_z} = n_b \ket{n_b, m_z}$ and $m_z$ is the z-projection of the collective operator $\hat{J}_z$ such that $\hat{J}_z \ket{n_b, m_z} = m_z \ket{n_b, m_z}$. Consider an initial state of the system $\ket{\mathrm{initial}}=\ket{n_b=0, m_z}$. The only intermediate states $\ket{\mathrm{intermediate}}$ such that $|\bra{\mathrm{intermediate}}\hat{V}\ket{\mathrm{initial}}|^2\neq 0$ are $\ket{\mathrm{intermediate}} = \ket{n_b=1, m_z\pm1}$. We will see such terms give rise to the exchange interactions $\hat{J}_+ \hat{J}_-$ and $\hat{J}_- \hat{J}_+$. At the same order in perturbation theory, there are additional non-zero matrix elements of the form $\bra{n_b=2, m_z=\pm2} \hat{V} \ket{\mathrm{n_b=1,m_z\pm 1}} \bra{n_b=1,m_z\pm 1}\hat{V} \ket{n_b=0,m_z}\neq0$. Such terms give rise to pair raising $\hat{J}^2_+$ and pair lowering $\hat{J}^2_-$ interactions. However these terms can be neglected since the initial and final states differ by $2\hbar \left( \Delta_d \pm  \omega_z\right)$ in energy. After identifying the intermediate states, we can re-write the Hamiltonian $\hat{H}_0$ in the $\ket{n_b,m_z}$ basis as
	\begin{equation}
	\begin{aligned}
	\hat{H}_0 =& \sum_{m_z} (\omega_z - \Delta_d) \ket{1,m_z+1}\bra{1,m_z+1} - (\omega_z + \Delta_d) \ket{1,m_z-1}\bra{1,m_z-1}\\
	& + \sum_{m_z} m_z \omega_z \left(\ket{1,m_z+1}\bra{1,m_z+1} + \ket{0,m_z}\bra{0,m_z} + \ket{1,m_z-1}\bra{1,m_z-1}\right) ,
	\end{aligned}    
	\end{equation}
	where the second line can be ignored as a constant for the perturbation theory. And the $\hat{V} =\hat{V}_- +\hat{V}_+ $ with
	\begin{equation}
	\hat{V}_+ = \sum_{m_z}\frac{g_0^2}{2\Delta_a} \alpha_0 \left[ \sqrt{\frac{N}{2} (\frac{N}{2}+1)-m_z(m_z+1)} \ket{1,m_z+1} \bra{0,m_z} +\sqrt{\frac{N}{2} (\frac{N}{2}+1)-m_z(m_z-1)} \ket{1,m_z-1} \bra{0,m_z} \right]
	\end{equation}
	where $\hat{V}_- = \hat{V}_+^\dagger$. Second-order perturbation theory including the damping gives rise to an non-Hermitian Hamiltonian
	\begin{equation}
	\begin{split}
	\hat{H}_{\mathrm{nh}} &= \hat{H}_0 - \frac{i}{2}  \hat{L}^\dagger \hat{L} \\
	&= \sum_{m_z}(\omega_z - \Delta_d-i\kappa/2) \ket{1,m_z+1} \bra{1,m_z+1} -(\omega_z + \Delta_d+i\kappa/2) \ket{1,m_z-1} \bra{1,m_z-1} ,    
	\end{split}
	\end{equation}
	from which we derive the effective Hamiltonian 
	\begin{equation}
	\begin{split}
	\hat{H}_{\mathrm{eff}} &= -\frac{1}{2} \hat{V}^- \left[ \hat{H}_{\mathrm{nh}}^{-1} + \left( \hat{H}_{\mathrm{nh}}^{-1}\right)^\dagger \right] \hat{V}^+\\
	& = \chi_+ \hat{J}_+ \hat{J}_- + \chi_- \hat{J}_- \hat{J}_+ =\underbrace{ (\chi_+ + \chi_-) }_{\chi} (\hat{J}^2 - \hat{J}_z^2)+
	(\chi_+ - \chi_-)  \hat{J}_z , 
	\label{eqn:Heff}
	\end{split}
	\end{equation}
	with the interaction strengths
	\begin{equation}
	\begin{split}
	\chi_+ &= \left(\frac{g_0^2}{4\Delta_a}\right)^2\frac{|\alpha_{d}|^2 \kappa_1}{\Delta_d^2+\kappa^2/4}  \frac{\Delta_d+\omega_z}{(\Delta_d+\omega_z)^2+\kappa^2/4}  , \\
	\chi_- &= \left(\frac{g_0^2}{4\Delta_a}\right)^2\frac{|\alpha_{d}|^2 \kappa_1}{\Delta_d^2+\kappa^2/4}  \frac{\Delta_d-\omega_z}{(\Delta_d-\omega_z)^2+\kappa^2/4}.
	\end{split}
	\end{equation}
	
	\noindent The effective jump operator describing the dissipation process is found as
	\begin{equation}
	\begin{split}
	\hat{L}_{\mathrm{eff}}&=\hat{L} \hat{H}_{\mathrm{nh}}^{-1} \hat{V}_{+} \\
	& = \sum_{m_z}\frac{g_0^2 \alpha_0 \sqrt{\kappa}}{4 \Delta_a} \left(\frac{\sqrt{\frac{N}{2} (\frac{N}{2}+1)- m(m+1)}}{\omega_z -\Delta_d - i\kappa/2} \left|0,m_z+1\right\rangle \left\langle 0,m_z\right| + \frac{\sqrt{\frac{N}{2} (\frac{N}{2}+1)- m(m-1)}}{-\omega_z - \Delta_d - i\kappa/2} \left|0,m_z-1\right\rangle \left\langle 0,m_z\right| \right) \\
	&= \frac{g_0^2 \alpha_0 \sqrt{\kappa}}{4 \Delta_a}  \left(\frac{1}{\omega_z -\Delta_d - i\kappa/2} \hat{J}_{+} + \frac{1}{-\omega_z -\Delta_d - i\kappa/2} \hat{J}_{-}\right) , 
	\end{split}
	\end{equation}
	which corresponds to the two superradiance processes where the dressing laser is $\pm\omega_z$ detuned from the dressed cavity resonance and the atoms change their momentum states with photons from the modulation sideband escaping from the cavity.
	
	Given the effective Hamiltonian and jump operators above, we can calculate Heisenberg equations of motion for the spin operators under mean-field approximation:
	\begin{equation}
	\begin{split}
	\left\langle \dot{\hat{J}}_{+}\right\rangle &=i\omega_{z}\hat{J}_{+}-i\left[2(\chi_{+}+\chi_{-})+i(\Gamma_{+}-\Gamma_{-})\right]\left\langle \hat{J}_{+}\right\rangle \left\langle \hat{J}_{z}\right\rangle , \\
	\left\langle \dot{\hat{J}}_{-}\right\rangle &=-i\omega_{z}\hat{J}_{-}+i\left[2(\chi_{+}+\chi_{-})-i(\Gamma_{+}-\Gamma_{-})\right]\left\langle \hat{J}_{-}\right\rangle \left\langle \hat{J}_{z}\right\rangle , \\
	\left\langle\dot{\hat{J}}_z\right\rangle&= -(\Gamma_+ - \Gamma_-) \left\langle \hat{J}_{+}\right\rangle \left\langle \hat{J}_{-}\right\rangle , \label{eq:effeom}
	\end{split}
	\end{equation}
	with $   \Gamma_\pm = |\alpha_0|^2\left(\frac{g_0^2}{4 \Delta_a}\right)^2 \frac{\kappa}{(\Delta_d\pm\omega_z)^2 + \kappa^2/4}$.

	\subsection{Mean-field dynamics for the full atom-cavity system}
	Here we give a different but more physical method to derive the system dynamics. Start from  full atom-cavity Hamiltonian Eq.~\eqref{eqn:atomcavity}, the equation of motions for the collective atomic operators is given by:
	\begin{equation}
	\begin{split}
	\left\langle \dot{\hat{J}}_{+}\right\rangle  & =i\omega_{z}\left\langle \hat{J}_{+}\right\rangle -i \frac{g_{0}^{2}}{2\Delta_{a}}\left\langle \hat{a}^{\dagger}\hat{a}\right\rangle \left\langle \hat{J}_{z}\right\rangle ,\\
	\left\langle \dot{\hat{J}}_{-}\right\rangle  & = -i\omega_{z}\left\langle \hat{J}_{-}\right\rangle +i \frac{g_{0}^{2}}{2\Delta_{a}}\left\langle \hat{a}^{\dagger}\hat{a}\right\rangle \left\langle \hat{J}_{z}\right\rangle ,\\
	\left\langle \hat{J}_{z}\right\rangle &=-i\frac{g_{0}^{2}}{4\Delta_{a}}\left\langle \hat{a}^{\dagger}\hat{a}\right\rangle \left(\left\langle \hat{J}_{+}\right\rangle -\left\langle \hat{J}_{-}\right\rangle \right).
	\label{eqn:eomspin}
	\end{split}
	\end{equation}
	
	\noindent We further define $\tilde{\hat{J}}_+=\hat{J}_+e^{-i\omega_z t}$ and $\tilde{\hat{J}}_-=\hat{J}_-e^{i\omega_z t}$, where $\tilde{\hat{J}}_{\pm}$ are slowly varying and thus assumed to be constants during the time scale set by $\Delta_d$ and $\omega_z$. Then the mean-field equations of motion for the cavity field operators are 
	\begin{equation}
	\begin{split}
	\frac{d}{dt}\left\langle \hat{a}\right\rangle  & =i(\Delta_d+i\kappa/2)\left\langle \hat{a}\right\rangle -i\alpha_{d} \sqrt{\kappa_1}-i\frac{g_0^{2}}{4\Delta_{a}} \left(\left\langle \tilde{\hat{J}}_+\right\rangle e^{i\omega_{z}t} +\left\langle \tilde{\hat{J}}_-\right\rangle e^{-i\omega_{z}t} \right) \left\langle \hat{a}\right\rangle  , \\
	\frac{d}{dt}\left\langle \hat{a}^{\dagger}\hat{a}\right\rangle  & =i \left(\alpha_d^{*}\sqrt{\kappa_1}\left\langle \hat{a}\right\rangle -\alpha_{d} \sqrt{\kappa_1}\left\langle \hat{a}^{\dagger}\right\rangle \right)-\kappa\left\langle \hat{a}^{\dagger}\hat{a}\right\rangle, 
	\label{eqn:eomaad}
	\end{split}
	\end{equation}
	where the atomic properties $\braket{\tilde {\hat{J}}_\pm}$ modulate the cavity field and thus create modulation sidebands that are crucial for the exchange interactions. The formal integration for the cavity field is given by
	\begin{equation}
	\left\langle \hat{a}\right\rangle _{t}=-i\alpha_{d}\sqrt{\kappa_{1}}\exp[i\int_{0}^{t}d\tau\ \mathcal{C}(\tau)]\int_{0}^{t}dt^{\prime}\ \exp[-i\int_{0}^{t^{\prime}}d\tau\ \mathcal{C}(\tau)],
	\end{equation}
	with
	\begin{equation}
	\begin{aligned}
	\mathcal{C}(t)&=\Delta_{d}+i\frac{\kappa}{2}-\frac{g_{0}^{2}}{4\Delta_{a}}\left(\left\langle \tilde{\hat{J}}_{+}\right\rangle e^{i\omega_{z}t}+\left\langle \tilde{\hat{J}}_{-}\right\rangle e^{-i\omega_{z}t}\right) 
	\\
	&=\Delta_{d}+i\frac{\kappa}{2} - \frac{g_{0}^{2}}{4\Delta_{a}}\left(\left\langle \tilde{\hat{J}}_{x}\right\rangle \cos\omega_{z}t-\left\langle \tilde{\hat{J}}_{y}\right\rangle \sin\omega_{z}t\right).  \\
	\end{aligned}
	\end{equation}

	\noindent We then have:
	\begin{equation}
	\begin{aligned}
	\exp\left[i\int_{0}^{t}d\tau\mathcal{C}(\tau)\right] & =e^{i\left(\Delta_{d}+i\kappa/2\right)t}\exp\left[i\frac{g_{0}^{2}\left\langle \tilde{\hat{J}}_{y}\right\rangle }{2\Delta_{a}\omega_{z}}\right]\exp\left[-i\frac{g_{0}^{2}}{2\Delta_{a}\omega_{z}}\left[\left\langle \tilde{\hat{J}}_{x}\right\rangle \sin\left(\omega_{z}t\right)+\left\langle \tilde{\hat{J}}_{y}\right\rangle \cos\left(\omega_{z}t\right)\right]\right]\\
	&=e^{i\left(\Delta_{d}+i\kappa/2\right)t}\exp\left[i\frac{g_{0}^{2}\left\langle \tilde{\hat{J}}_{y}\right\rangle }{2\Delta_{a}\omega_{z}}\right]\sum_{n=-\infty}^{+\infty}(-1)^{n}J_{n}\left(\frac{g_{0}^{2}\left\langle \tilde{\hat{J}}_{x}\right\rangle }{2\Delta_{0}\omega_{z}}\right)e^{in\omega_{z}t}\sum_{m=-\infty}^{+\infty}(-i)^{m}J_{m}\left(\frac{g_{0}^{2}\left\langle \tilde{\hat{J}}_{y}\right\rangle }{\Delta_{0}\omega_{z}}\right)e^{im\omega_{z}t}\\
	&=e^{i\left(\Delta_{d}+i\kappa/2\right)t}\exp\left[i\frac{g_{0}^{2}\left\langle \tilde{\hat{J}}_{y}\right\rangle }{2\Delta_{a}\omega_{z}}\right]\sum_{n=-\infty}^{+\infty}(-1)^{n}J_{n}\left(\frac{g_{0}^{2}\left|\left\langle \tilde{\hat{J}}_{+}\right\rangle \right|}{2\Delta_{a}\omega_{z}}\right)e^{in\theta}e^{in\omega_{z}t}\\
	&=e^{i\left(\Delta_{d}+i\kappa/2\right)t} e^{i\Theta} ,
	\end{aligned}
	\end{equation}
	where $\tan\theta = \left\langle \tilde{\hat{J}}_{y}\right\rangle / \left\langle \tilde{\hat{J}}_{x}\right\rangle$, and $J_n(z)$ is the Bessel function of the first kind. In the derivation above, we use the Bessel function properties:
	\begin{equation}
	e^{i z \cos \phi}=\sum_{n=-\infty}^{+\infty} i^n J_n(z) e^{i n \phi}, \quad e^{i z \sin \phi}=\sum_{n=-\infty}^{+\infty} J_n(z) e^{i n \phi},
	\end{equation}
	as well as 
	\begin{equation}
	e^{i \nu \phi} J_\nu(R)=\sum_{k=-\infty}^{\infty} J_k(\rho) J_{\nu+k}(r) e^{i k \varphi}, \quad R=\sqrt{r^2+\rho^2-2 r \rho \cos \varphi}, \quad e^{2 i \phi}=\frac{r-\rho e^{-i \varphi}}{r-\rho e^{i \varphi}} .
	\end{equation}
	
	\noindent Similarly, we have:
	\begin{equation}
	\int_{0}^{t}dt^{\prime}\exp\left[-i\int_{0}^{t^{\prime}}d\tau\mathcal{C}(\tau)\right]=i\exp\left[-i\frac{g_{c}^{2}\left\langle \tilde{\hat{J}}_{y}\right\rangle }{\Delta_{0}\omega_{z}}\right]\sum_{n=-\infty}^{+\infty}(-1)^{n}J_{n}\left(\frac{g_{c}^{2}\left|\left\langle \tilde{\hat{J}}_{+}\right\rangle \right|}{\Delta_{0}\omega_{z}}\right)e^{-in\theta}\frac{e^{-i\left(\Delta_{d}+i\kappa/2+n\omega_{z}\right)t}-1}{\Delta_{d}+i\kappa/2+n\omega_{z}} .
	\end{equation}
	
	In the limit $g_{c}^{2}N/4\Delta_{0}\ll\omega_{z}$, we expand to the first order of the Bessel function and solve the cavity field at long time with $t\gg 1/\kappa$
	\begin{equation}
	\left\langle \hat{a}\right\rangle _{t}\approx\frac{\alpha_{d}\sqrt{\kappa_{1}}}{\Delta_{d}+i\kappa/2}e^{i\Theta} \left[1+\frac{g_{0}^{2}\left\langle \tilde{\hat{J}}^{+}\right\rangle}{4\Delta_{a}\omega_{z}}\frac{\Delta_{d}+i\kappa/2}{\Delta_{d}-\omega_{z}+i\kappa/2} e^{i\omega_{z}t}-\frac{g_{0}^{2}\left\langle \tilde{\hat{J}}^{-}\right\rangle}{4\Delta_{a}\omega_{z}}\frac{\Delta_{d}+i\kappa/2}{\Delta_{d}+\omega_{z}+i\kappa/2} e^{-i\omega_{z}t}\right] .
	\label{eqn:cavityfieldat}
	\end{equation}
	
	From Eq. \eqref{eqn:eomaad}, the formal integration for $\left\langle \hat{a}^{\dagger}\hat{a}\right\rangle $
	is given by,
	\begin{equation}
	\left\langle \hat{a}^{\dagger}\hat{a}\right\rangle _{t}=e^{-\kappa t}\int_{0}^{t}d\tau\ ie^{\kappa\tau}\sqrt{\kappa_{1}}\left(\alpha_{d}^{*}\left\langle \hat{a}\right\rangle _{\tau}-\alpha_{d}\left\langle \hat{a}^{\dagger}\right\rangle _{\tau}\right).
	\end{equation}
	With Eq.~\eqref{eqn:cavityfieldat}, we have:
	\begin{equation}
	\begin{aligned}
	\left\langle \hat{a}^{\dagger}\hat{a}\right\rangle _{t}&=\left|\alpha_{0}\right|^{2}+\frac{g_{0}^{2}|\alpha_{d}|^{2}\kappa_{1}}{4\Delta_{a}\omega_{z}}\left[\frac{1}{\left(\Delta_{d}-\omega_{z}+i\kappa/2\right)\left(\Delta_{d}-i\kappa/2\right)}-\frac{1}{\left(\Delta_{d}+\omega_{z}-i\kappa/2\right)\left(\Delta_{d}+i\kappa/2\right)}\right]\left\langle \tilde{\hat{J}}_{+}\right\rangle e^{i\omega_{z}t} \\
	&+\frac{g_{0}^{2}|\alpha_{d}|^{2}\kappa_{1}}{4\Delta_{a}\omega_{z}}\left[\frac{1}{\left(\Delta_{d}-\omega_{z}-i\kappa/2\right)\left(\Delta_{d}+i\kappa/2\right)}-\frac{1}{\left(\Delta_{d}+\omega_{z}+i\kappa/2\right)\left(\Delta_{d}-i\kappa/2\right)}\right]\left\langle \tilde{\hat{J}}_{-}\right\rangle e^{-i\omega_{z}t} \\
	& \equiv \left|\alpha_{0}\right|^{2}+A\left\langle \tilde{\hat{J}}_{+}\right\rangle e^{i\omega_{z}t}+A^{*}\left\langle \tilde{\hat{J}}_{-}\right\rangle e^{-i\omega_{z}t}.
	\end{aligned}
	\end{equation}
	Substituting the solution $\left\langle \hat{a}^{\dagger}\hat{a}\right\rangle_{t}$ in Eq.~\eqref{eqn:eomspin}, here are the equations of motion with only atom-atom interaction
	\begin{equation}
	\begin{aligned}
	\left\langle \dot{\hat{J}}_{+}\right\rangle &=i\omega_{z}\left\langle \hat{J}_{+}\right\rangle -i\frac{g_{0}^{2}}{2\Delta_{a}}\left(\left|\alpha_{0}\right|^{2}+A\left\langle \tilde{\hat{J}}_{+}\right\rangle e^{i\omega_{z}t}+A^{*}\left\langle \tilde{\hat{J}}_{-}\right\rangle e^{-i\omega_{z}t}\right)\left\langle \hat{J}_{z}\right\rangle ,\\
	\left\langle \dot{\hat{J}}_{-}\right\rangle &=-i\omega_{z}\left\langle \hat{J}_{-}\right\rangle +i\frac{g_{0}^{2}}{2\Delta_{a}}\left(\left|\alpha_{0}\right|^{2}+A\left\langle \tilde{\hat{J}}_{+}\right\rangle e^{i\omega_{z}t}+A^{*}\left\langle \tilde{\hat{J}}_{-}\right\rangle e^{-i\omega_{z}t}\right)\left\langle \hat{J}_{z}\right\rangle ,\\
	\left\langle \dot{\hat{J}}_{z}\right\rangle  &=-i\frac{g_{0}^{2}}{4\Delta_{a}}\left(\left|\alpha_{0}\right|^{2}+A\left\langle \tilde{\hat{J}}_{+}\right\rangle e^{i\omega_{z}t}+A^{*}\left\langle \tilde{\hat{J}}_{-}\right\rangle e^{-i\omega_{z}t}\right)\left(\left\langle \tilde{\hat{J}}_{+}\right\rangle e^{i\omega_{z}t}-\left\langle \tilde{\hat{J}}_{-}\right\rangle e^{-i\omega_{z}t}\right).
	\end{aligned}
	\end{equation}
	
	With the slowly varying  variables $\left\langle \tilde{\hat{J}}_{\pm}\right\rangle $, the above equation of motion can be further simplified:
	\begin{equation}
	\begin{aligned}
	\left\langle \dot{\tilde{\hat{J}}}_{+}\right\rangle &=-i\frac{g_{0}^{2}}{2\Delta_{a}}A\left\langle \tilde{\hat{J}}_{+}\right\rangle \left\langle \hat{J}_{z}\right\rangle ,\\
	\left\langle \dot{\tilde{\hat{J}}}_{-}\right\rangle &=i\frac{g_{0}^{2}}{2\Delta_{a}}A^{*}\left\langle \tilde{\hat{J}}_{-}\right\rangle \left\langle \hat{J}_{z}\right\rangle ,\\
	\left\langle \dot{\hat{J}}_{z}\right\rangle  &=-\frac{g_{0}^{2}}{2\Delta_{a}}\frac{A-A^{*}}{2i}\left\langle \tilde{\hat{J}}_{+}\right\rangle \left\langle \tilde{\hat{J}}_{-}\right\rangle ,
	\label{eqn:eomsimple}
	\end{aligned}
	\end{equation}
	
	\noindent where the single-particle processes (i.e. off-resonant Bragg coupling) and the fast oscillating dynamics at frequency $2\omega_z$ are ignored. With $g_0^2 A / 2\Delta_0=2(\chi_+ + \chi_-) + i (\Gamma_+ - \Gamma_-)$, the above equations of motion agree with Eq.~\eqref{eq:effeom} and thus verify the predicted dynamics from the effective spin Hamiltonian. In the limit without superradiance, $\left\langle \dot{\hat{J}}_{z}\right\rangle=0$ such that the total spin projection is conserved.
	
	\subsection{Exchange interaction versus one-axis twisting}
	In the absence of inhomogeneity, the exchange interaction predicts the same dynamics for the collective operators as the one-axis twisting interaction. To emphasize the contrast between the two, we now consider a minimal model for two spin ensembles  $\hat{J}_1$ and $\hat{J}_2$ with a differential transition frequency $\delta$, in light of Fig.~4(C) with two spin ensembles defined for $p>0$ and $p<0$. 
	
	For the exchange interaction with the Hamiltonian $\hat{H}=\hbar \chi \left(\hat{J}_{1+} + \hat{J}_{2+} \right) \left(\hat{J}_{1-} + \hat{J}_{2-}\right)+ \frac{\hbar \delta}{2} \left( \hat{J}_{1z}-\hat{J}_{2z} \right)$, here are the equations of motion for individual spin operators
	\begin{equation}
	\begin{aligned}
	\braket{ \dot{ \hat{J}}_{1+}} &=-i2\chi \left( \braket{\hat{J}_{1+}}+ \braket{\hat{J}_{2+}} \right)  \braket{\hat{J}_{1z}}+i\delta  \braket{\hat{J}_{1+}}, \\
	\braket{ \dot{\hat{J}}_{2+}} &=-i2\chi \left( \braket{\hat{J}_{1+}}+ \braket{\hat{J}_{2+}} \right)  \braket{\hat{J}_{2z}}-i\delta  \braket{\hat{J}_{2+}}, \\
	\braket{ \dot{\hat{J}}_{1z}}&=i\chi\left(  \braket{\hat{J}_{1-}}  \braket{\hat{J}_{2+}}-  \braket{\hat{J}_{1+}}  \braket{\hat{J}_{2-}}\right),\\ 
	\braket{\dot{\hat{J}}_{2z}}&=i\chi\left( \braket{\hat{J}_{1+}}  \braket{\hat{J}_{2-}}-  \braket{\hat{J}_{1-}}  \braket{\hat{J}_{2+}}\right),
	\end{aligned}
	\end{equation}
	from which we can clearly see that the inhomogenity term causes oscillations for the coherence $\braket{\hat{J}_{1+}}$ and $\braket{\hat{J}_{2+}}$ and population differences $\braket{\hat{J}_{1z}}$ and $\braket{\hat{J}_{2z}}$ as discussed in Fig.~4(C). Also, with the exchange interaction, the total spin projection $\braket{\hat{J}_z}$ converses given $\frac{d \braket{\hat{J}_z}}{dt}=\frac{d \braket{\hat{J}_{1z}}}{dt}+\frac{d \braket{\hat{J}_{2z}}}{dt}=0$, while the individual spin projections $\braket{\hat{J}_{1z}}$ and $\braket{\hat{J}_{2z}}$ are not conserved.
	
	In contrast, the equations of motion for OAT with inhomogenity $\hat{H}=\hbar \chi \left( \hat{J}_{1z}+\hat{J}_{2z} \right)^2+\frac{\hbar\delta}{2}\left( \hat{J}_{1z}-\hat{J}_{2z} \right)$ are
	\begin{equation}
	\begin{aligned}
	\braket{ \dot{\hat{J}}_{1+}} & =-i 2 \chi \left( \braket{\hat{J}_{1+}} + \braket{\hat{J}_{2+}}\right) \braket{\hat{J}_{1z}} ,\\
	\braket{ \dot{\hat{J}}_{2+}} & =-i 2 \chi \left( \braket{\hat{J}_{1+}} + \braket{\hat{J}_{2+}} \right) \braket{\hat{J}_{2z}}.
	\end{aligned}
	\end{equation}
	Because the inhomogenity term commutes with the interaction term, both the total and individual spin projections $\braket{\hat{J}_z}$, $\braket{\hat{J}_{1z}}$ and $\braket{\hat{J}_{2z}}$ are conserved.
	
	For all the simulation results shown in the main text, we account for the effect of Doppler broadening (inhomogeneous kinetic energy difference) by including $\hat{H}_{\mathrm{in}}$ for all the momentum states in the equations of motion and solving the equations of motion numerically.

	\section{Atomic probe, Bragg and dressing laser}
	A single laser coupled into the cavity is used to realize the quantum non-demolition (QND) measurement of the momentum state populations, to create the dressing laser for momentum exchange interactions, and to realize Bragg rotations.  We refer to this laser generically as the atomic probe, and we realize these various functions via rapid control of its frequency and phase as described below.  In all cases, the relevant cavity resonance is stabilized to the blue of the atomic transition frequency $\omega_{a}$ with detuning $\Delta_a\approx 2\pi \times 500$~MHz.
	
	For the QND measurement we measure a time-averaged shift of the cavity resonance frequency that depends approximately linearly on the number of atoms in the $\ket{F=2, m_F=0}$ state. We do this by creating a probe tone incident on the cavity that is swept across resonance in time.  We then detect this reflected light using homodyne detection.  To achieve this, the atomic probe laser is  blue detuned of the cavity by  nominally $80$~MHz.  The homodyne reference beam or local oscillator is created by picking off a fraction of the laser light and shifting it to the red by 80~MHz using an acousto-optic modulator (AOM). The probe tone is generated using a fiber phase modulator weakly driven with 80~MHz.  The lower phase modulation sideband serves as the actual tone used to measure the cavity frequency shift and is at the same frequency as (and phase coherent with) the homodyne reference laser.   The much stronger carrier is nominally $80~$MHz from the cavity resonance and therefore simply reflects off of the cavity and generates a heterodyne tone at $80$~MHz in the homodyne detector.  We detect the phase of this 80~MHz signal to monitor optical path length changes and then actively stabilize the homodyne detection quadrature via feeding back on the phase of the 80~MHz frequency sent to the AOM used to generate the homodyne reference beam. The cavity frequency shift is measured by sweeping the atomic probe laser's frequency over a range of 1~MHz from below to above the cavity resonance frequency in approximately 200~$\mu$s and fitting the observed dispersive signal from the homodyne detector.
	
	
	Driving the Bragg rotations between $\ket{p_0-\hbar k}$ and $\ket{p_0+\hbar k}$ requires two different optical tones on the atomic probe beam separated by $\omega_z$. We realize this by first red shift the atomic probe by 75~MHz with one AOM and then blue shift it back with another AOM driven with two radio frequency (RF) tones at 75~MHz and $(75-\omega_z/2\pi)$~MHz before sending it into the fiber phase modulator. Because the atoms accelerate under gravity, we linearly chirp the frequency separation $\omega_z/2\pi$ between the two sidebands by $25.11$~kHz/ms to compensate the changing Doppler shift.  This light is then sent through the same fiber phase modulator as above before being injected into the cavity. By stabilizing the atomic probe carrier at 77~MHz from the cavity, the lower phase modulation sidebands now consists of two tones separated by $\omega_z$ that are 3 MHz detuned from the cavity resonance.  These tones non-resonantly enter the cavity and drive the Bragg rotations. With the detuning $3~\mathrm{MHz} \gg \kappa$, we suppress any frequency noise to amplitude or phase noise conversion that would degrade the fidelity of the Bragg rotations. For all the Bragg pulses applied in the interferometer sequence, the Rabi frequency is $8.3~$kHz, giving a $\pi$-pulse duration of $60.~\mu$s.

	For driving the momentum exchange interaction, we apply only one RF tone at 75~MHz on the (blue shifting) AOM before the phase modulation as for the QND measurement and fast switch the RF frequency for driving the fiber phase modulator to put the atomic probe tone about few hundreds kHz detuned from the dressed resonance.

	\section{Initial state preparation and readout}
	The atoms are initially prepared via velocity-dependent Raman transitions in a very narrow momentum range and hyperfine state given by$\ket{F=2,m_F=0,p_0-\hbar k}$. This specific hyperfine state is favorable for the velocity-selection since using the Rb clock states avoids broadening of the transition due to magnetic fields.  However, to exploit the more favorable Clebsch-Gordan coefficients of the $\ket{F=2,m_F=2}$ to excited $\ket{F'=3,m_F=3}$ cycling transition, we use a series of microwave $\pi$-pulses to transfer the atoms into $\ket{F=2,m_F=2}$ without significantly shifting their momentum states. We first apply a microwave $\pi$-pulse to transfer atoms into $\ket{F=1,mF=1,p_0-\hbar k}$, blow away any untransferred atoms with a laser beam on resonance with $\ket{F=2}\rightarrow \ket{F'=3}$ transition then a second microwave $\pi$-pulse transfer the atoms into $\ket{F=2,m_F=2,p_0-\hbar k}$. This configuration is the initial state from which we realize the Bragg interferometer and momentum exchange interactions. 
	
	To read out the populations in individual momentum states at the end of the experimental sequences, we begin by using microwave $\pi$-pulses to map the population back from $\ket{F=2, m_F=2}$ to $\ket{F=1, m_F=0}$.  The process is the reverse of the above, but with an added $\pi$-pulse at the end to transfer atoms from $\ket{F=2, m_F=0}$ to $\ket{F=1, m_F=0}$.We then apply a Raman $\pi$-pulse (two-photon Rabi frequency $4.2~$kHz) for transferring atoms from $\ket{F=1,m_F=0,p_0-\hbar k}$ to $\ket{F=2,m_F=0,p+\hbar k}$. A QND measurement is then used to measure the number of atoms or population as described above. We then blow away the $\ket{F=2}$ atoms with a laser beam resonant with the atomic $\ket{F=2\rightarrow 3'}$ transition.  We then measure the population in $p_0+\hbar k$ by applying a Raman $\pi$-pulse again with the appropriate two-photon detuning and perform a second QND measurement.  The momentum exchange interaction's residual superradiance may transfer atoms into adjacent momentum states $p_0\pm 3 \hbar k$. We iterate the above procedure to measure these populations as well. Thus, on every run of the experiment, we measure the populations in the four momentum states $p_0\pm \hbar k$ and $p_0\pm 3\hbar k$.
	

	

	\section{Author contributions}
	
	C.L. and V.P.W.K. contributed to the building of the experiment, conducted the experiments and data analysis. J.K.T. conceived and supervised the experiments. H.Z., J.D.W. and A.C. contributed to the theoretical derivation and numerical simulations supervised by M.J.H. and A.M.R.  C.L. and J.K.T. wrote the manuscript with feedback from A.M.R. All authors discussed the experiment implementation and results and contributed to the manuscript.
	
	\section{Data availability}
	
	All data obtained in the study are available from the corresponding author upon reasonable request.
	
	\section{Competing interests}
	
	The authors declare no competing interests.

	\clearpage

\end{document}